\documentclass[11pt]{article}

\usepackage[skip=14pt plus2pt, indent=0pt]{parskip}
\usepackage{verbatim}
\usepackage{epsfig,graphics}
\usepackage {graphicx}
\usepackage {epsfig}
\usepackage {subfigure}
\usepackage {tabularx} 
\usepackage{rotate}	
\usepackage{slashed}
\usepackage{enumerate}
\usepackage{amsmath}

\usepackage{amsfonts}
\usepackage{amssymb}
\usepackage{graphicx}
\usepackage{xcolor}
\usepackage{cite}
\usepackage{mathrsfs}

\usepackage{fancyhdr}
\usepackage[hidelinks]{hyperref}

\usepackage{braket}
\usepackage{cancel}

\def\beq{\begin{equation}}
\def\eeq{\end{equation}}
\def\beqa{\begin{eqnarray}}
\def\eeqa{\end{eqnarray}}

\def\Mpd{M_{\text{pl,}\, d}}
\def\Mpf{M_{\text{pl,}\, 4}}
\def\Mpt{M_{\text{pl,}\,  10}}

\def\LQG{\Lambda_{\text{QG}}}

\catcode`\@=11   
\newdimen\@rotdimen
\newbox\@rotbox  

\def\@vspec#1{\special{ps:#1}}
\def\@rotstart#1{\@vspec{gsave currentpoint currentpoint translate
		#1 neg exch neg exch translate}}
\def\@rotfinish{\@vspec{currentpoint grestore moveto}}
\def\@rotr#1{\@rotdimen=\ht#1\advance\@rotdimen by\dp#1%
	\hbox to\@rotdimen{\hskip\ht#1\vbox to\wd#1{\@rotstart{90 rotate}%
			\box#1\vss}\hss}\@rotfinish}
\def\@rotl#1{\@rotdimen=\ht#1\advance\@rotdimen by\dp#1%
	\hbox to\@rotdimen{\vbox to\wd#1{\vskip\wd#1\@rotstart{270 rotate}%
			\box#1\vss}\hss}\@rotfinish}%
\def\@rotu#1{\@rotdimen=\ht#1\advance\@rotdimen by\dp#1%
	\hbox to\wd#1{\hskip\wd#1\vbox to\@rotdimen{\vskip\@rotdimen
			\@rotstart{-1 dup scale}\box#1\vss}\hss}\@rotfinish}%
\def\@rotf#1{\hbox to\wd#1{\hskip\wd#1\@rotstart{-1 1 scale}%
		\box#1\hss}\@rotfinish}%
\def\rotate{\@ifnextchar[{\@rotate}{\@rotate[l]}}
\def\@rotate[#1]#2{\setbox\@rotbox=\hbox{#2}\@nameuse{@rot#1}\@rotbox}

\catcode`\@=12

\topmargin
-1.5cm
\textwidth
15.5cm
\textheight
23.5cm
\oddsidemargin
0.7cm
\evensidemargin
0.7cm

\begin{document}
\makeatletter
\@addtoreset{equation}{section}
\makeatother
\renewcommand{\theequation}{\thesection.\arabic{equation}}
\pagestyle{empty}
\vspace{-0.2cm}
\rightline{IFT-UAM/CSIC-23-131}
\rightline{MPP-2023-419}
\vspace{1.2cm}
\begin{center}
		
		\LARGE{ On the Species Scale, Modular Invariance \\and the Gravitational EFT expansion \\
			[13mm]}
		
		\large{ A. Castellano$^\clubsuit$, A. Herr\'aez$^\diamondsuit$ and L.E. Ib\'a\~nez$^\clubsuit$
			\\[12mm]}
		\small{
			$^\clubsuit$ {Departamento de F\'{\i}sica Te\'orica
				and Instituto de F\'{\i}sica Te\'orica UAM/CSIC,\\
				Universidad Aut\'onoma de Madrid,
				Cantoblanco, 28049 Madrid, Spain}  \\[5pt]
			$^\diamondsuit$  {Max-Planck-Institut f\"ur Physik (Werner-Heisenberg-Institut),\\ F\"ohringer Ring 6, 80805 M\"unchen, Germany} 
			\\[20mm]
		}
		\small{\bf Abstract} \\[6mm]
	\end{center}
	\begin{center}
		\begin{minipage}[h]{15.60cm}
The concept of the species scale as the quantum gravity cut-off has been recently emphasised in the context of the Swampland program. Along these lines, we continue the quest for a precise understanding of its role within effective field theories of gravity as well as a global definition of the latter in case there is enough supersymmetry preserved. To do so, we exploit duality symmetries, the familiar asymptotic dependence imposed by the presence of infinite towers of light states and the known behaviour of higher-curvature corrections to the Einstein-Hilbert action in various String Theory setups. In those cases, we obtain a self-consistent result for the identification of the 
species scale as the quantum gravity cut-off, but also present some puzzles related to the suppression of certain higher-dimensional operators as well as minor ambiguities that may arise in the deep interior of moduli space. 
		\end{minipage}
	\end{center}
	\newpage
\pagestyle{empty}
\renewcommand{\thefootnote}{\arabic{footnote}}
\setcounter{footnote}{0}


	
\tableofcontents
	
\pagestyle{empty}
\newpage
\setcounter{page}{1}
\pagestyle{plain}

\section{Introduction}
\label{sec:intro}

The goal of the Swampland program \cite{Vafa:2005ui} (see also \cite{Brennan:2017rbf,Palti:2019pca,vanBeest:2021lhn,Grana:2021zvf,Harlow:2022ich,Agmon:2022thq,VanRiet:2023pnx} for reviews) is to distinguish the set of effective field theories (EFTs) that can be consistently coupled to quantum gravity (QG) from those which cannot. In particular, the main focus is put on gravitational EFTs that reduce to Einstein gravity coupled to matter in the infra-red (IR). However, the fact that they are merely effective descriptions can be translated into the existence of generically infinite towers of higher-dimensional operators that are suppressed at low energies. Thus, one expects the following schematic form of the gravitational EFT action in $d$ spacetime dimensions to appear
\beq
S_{\mathrm{EFT,\,}d}=\int d^d x \, \sqrt{-g}\ \dfrac{1}{2\kappa_d^2}\left(R + \sum_n \frac{\mathcal{O}_n (R)}{\LQG^{n-2}}\right)+\ldots \, ,
\label{eq:gravEFTexpansionI}
\eeq
where $\mathcal{O}_n (R)$ represent higher-curvature operators of mass dimension $n>2$ and the ellipsis accounts for any possible matter that may also be present in the theory. Since all these corrections can be generated (at least) by quantum effects, the natural question that arises is at which energy scale they start kicking in, or equivalently, when do QG phenomena become important. Surprisingly, even though one may think --- based on simple dimensional analysis --- that this scale is generically set by the Planck mass, this does not actually seem to be the whole story in quantum gravity (see e.g., \cite{Donoghue:1994dn,Donoghue:2022eay,Donoghue:1995cz,Burgess:2003jk,Alvarez:2020zul} for a review on EFT approaches to gravitational theories and relevant scales therein).

From another point of view, it is known that one of the most thoroughly tested Swampland conditions, namely the Distance Conjecture \cite{Ooguri:2006in}, predicts the existence of infinite towers of states becoming light at infinite distance boundaries within the moduli spaces of quantum-gravitational theories, and thus a corresponding decrease in the cut-off of the EFT. These two perspectives resonate with each other and converge beautifully once one realises that the QG cut-off is actually not given by the naive guess from simple dimensional analysis, namely the Planck scale, but should be rather identified with the species scale \cite{Dvali:2007hz,Dvali:2007wp, Dvali:2009ks,Dvali:2010vm}. The key ingredient incorporated in the original proposal of the species scale is precisely the idea that an arbitrary number of light, weakly coupled degrees of freedom, $N$, may actually lower the quantum gravity cut-off in $d$-dimensions as follows
\begin{equation}\label{eq:Lambdaspdef}
    \Lambda_{\text{QG}}=\dfrac{M_{\text{pl},\, d}}{N^{\frac{1}{d-2}}}\, .
\end{equation}
This proposal, which was motivated from both perturbative and non-perturbative arguments (see e.g., \cite{Castellano:2021mmx} and references therein) 
has been the subject of exhaustive studies in recent times due to its interesting connection with the Swampland program, as originally highlighted in \cite{Grimm:2018ohb} (see also \cite{Corvilain:2018lgw,Castellano:2021mmx,Castellano:2022bvr,Castellano:2023qhp,Calderon-Infante:2023ler,vandeHeisteeg:2022btw,Cribiori:2023ffn,Cribiori:2023sch,Blumenhagen:2023yws,Blumenhagen:2023tev,Blumenhagen:2023xmk,vandeHeisteeg:2023ubh,Andriot:2023isc,Cota:2022yjw,Cota:2022maf,Calderon-Infante:2023uhz} for some recent applications related to the Distance Conjecture, the Weak Gravity Conjecture \cite{Arkani-Hamed:2006emk} and their interplay with Black Hole entropy and the Emergence Proposal). In particular, it is crucial to emphasise that the species scale encodes more information than that captured by accounting only for the lightest tower of states, as predicted by the Distance Conjecture, since other (even parametrically) heavier towers can also contribute significantly to the number of species \cite{Castellano:2021mmx} and thus be relevant for the dual descriptions emerging as one approaches the infinite distance points.

However, even though the species scale is by now fairly well understood close to weak coupling points arising at infinite distance limits --- where the notion of light species is usually well-defined, it remains to be elucidated how (if at all possible) to make sense out of it as a scale defined over all moduli space. In particular, this avenue has been pursued for the first time in \cite{vandeHeisteeg:2022btw} (see also \cite{Cribiori:2023sch}), where the authors emphasised the role of \eqref{eq:Lambdaspdef} as the true quantum gravity cut-off from its purported property as being the scale suppressing the higher-curvature corrections to the low energy Einstein-Hilbert action, even in the deep interior of moduli space. In a specific 4d $\mathcal{N}=2$ setup arising from Type II Calabi--Yau compactifications, where a set of such corrections (in particular the $R^2$-term) can be computed \emph{exactly}, the authors proposed some general definition of the species scale that reproduced all known asymptotic behaviours, as well as predicted the location of the desert point \cite{Long:2021jlv}, namely the locus in moduli space where $\Lambda_{\text{QG}}$ can be pushed to its maximum. The goal of this work is to continue along these lines, thus trying to study systematically whether the species cut-off is indeed the relevant energy scale controlling the EFT expansion in gravity. To do so, we analyze several supersymmetric examples in the String Theory literature where the first low-lying gravitational corrections to the two-derivative action are known \cite{Green:2010kv,Green:2010wi}, and find perfect agreement with our expectations. We also briefly consider the possibility of whether duality symmetries as well as \emph{appropriate} asymptotic behaviour alone are enough so as to determine unequivocally the global form of the species scale, analysing in detail the $SL(2,\mathbb{Z})$ case as a proof of concept. Notably, even though we cannot single out such globally defined function, we are able to provide some candidate that correctly captures its main qualitative features, such as the correct asymptotic moduli dependence or the presence of a desert point.

Along the way, we are faced with several questions that we believe are relevant for a complete understanding of the role of the species scale in EFTs of gravity in general. In particular, our results show that the species cut-off seems to control the first non-vanishing curvature corrections in the setups under considerations, i.e. the $R^2$-operator in 4d $\mathcal{N}=2$ or the $R^4$-term in their 8d, 9d and 10d higher supersymmetric counterparts. These may be compared with other higher-dimensional operators computed in the literature, such as the ones that include powers of the graviphoton field strength in four dimensions (e.g., $R^2 F^{2k}$ for $k\geq 1$) or those involving additional spacetime derivatives of the aforementioned curvature invariants in $d=8,9,10$ (e.g., $\partial^4 R^4$ or $\partial^6 R^4$). Interestingly, we find that they are not always suppressed by the species cut-off, but rather by the mass of the lightest state in the leading infinite tower. This opens up some possibilities, such as the option that it is the relevant/marginal gravitational operators (in the Wilsonian sense) which are generically suppressed by the quantum gravity cut-off, whilst the irrelevant ones could be controlled instead by other lighter scales. Furthermore, even when both scales coincide (e.g., when the quantum gravity cut-off is set by the string scale) one obtains that there might exist some ambiguity in the global definition of $\LQG$ from focusing just on the first correction, since other higher-derivatice terms are suppressed by modular functions which cannot be expressed as powers of one another. Nonetheless, we find that this ambiguity actually disappears in the asymptotic limits --- where a weak coupling expansion arises --- and we moreover expect it to not modify certain qualitative features such as the presence or location of a desert point \cite{Long:2021jlv}. Finally, we also find an intriguing connection between the definition of the species scale and the eigenfunctions of certain elliptic operators constructed out of the moduli space metric (once any present discrete duality is modded out), and speculate on the possibility that this might also be a defining property of any such `global' species scale function.

The paper is organised as follows. In Section \ref{s:speciesscale} we provide a brief review of the status of the species scale and study the predictive power of dualities regarding the determination of a globally defined species function. In Section \ref{s:Examples} we analyze certain higher-curvature corrections which are known to arise in stringy constructions with maximal supersymmetry in 10d, 9d and 8d, paying special attention to the energy scale suppressing those. We comment and extend in Section \ref{s:4dN=2} the 4d 
$\mathcal{N}=2$ setup studied in \cite{vandeHeisteeg:2022btw}, elaborating on the role of other higher-curvature corrections, and presenting some general lessons on how the gravitational EFT seems to behave with respect to the aforementioned scales. We leave some general comments and discussions for Section \ref{s:discussion}, and relegate some background material on the relevant automorphic forms and the maximal supergravity action in 8d for Appendices \ref{ap:Massform} and \ref{ap:maxSUGRA8d}, respectively.

\section{The Species Scale}
\label{s:speciesscale}

\subsection{The Basic Idea}
\label{ss:basics}
The main goal of this work is to understand the role and provide a proper definition of the species scale, understood as the scale at which quantum gravitational effects can no longer be neglected, all over the moduli space. To do so, one can try to approach the problem from several different angles. The first one is to try to define this EFT cut-off as the scale at which corrections to Einstein gravity become important. In general, one expects the lagrangian density of such gravitational EFT to be organised according to the following energy expansion
\beq
\mathcal{L}_{\mathrm{EFT,\,}d}\, =\, \dfrac{1}{2\kappa_d^2}\left(R + \sum_n \frac{\mathcal{O}_n (R)}{\LQG^{n-2}(\phi)}\right) + \frac{1}{2} G_{ij} \partial_{\mu} \phi^i \partial^{\mu} \phi^j+\ldots \, ,
\label{eq:gravEFTexpansion}
\eeq
with $\kappa_d^{2}=M_{\text{pl,\,}d}^{2-d}$ being Einstein's gravitational coupling, $\mathcal{O}_n (R)$ representing any dimension-$n$ higher-curvature correction (e.g., $R^2$ for $n=4$ or $R^4$ for $n=8$) and $\Lambda_{\text{QG}}(\phi)$ giving precisely the energy scale controlling/suppressing such corrections, which moreover generically depends on the massless scalar field vevs $\braket{\phi^j}$. The ellipsis indicates any other couplings and matter present in the EFT. Given the non-renormalizability of Einstein's theory, the naive expectation from simple dimensional analys is for such ultra-violet (UV) scale to be precisely around the Planck scale. However, in the presence of $N$ light weakly coupled species, this UV cut-off can be significantly lowered, and typically coincides with the so-called species scale \cite{Dvali:2007hz,Dvali:2007wp}
\begin{equation}
\label{eq:speciesscale}
    \Lambda_{\text{sp}}\simeq \dfrac{M_{\text{pl,\,}d}}{N^{\frac{1}{d-2}}}\, .
\end{equation}
This can be seen from e.g., computing the one-loop contribution from such species to the 2-point function of the graviton and observing that the one-loop correction and the tree-level piece become of the same order at precisely the scale defined in \eqref{eq:speciesscale}.

An alternative, and perhaps more robust approach than the aforementioned perturbative computation, is to consider instead the smallest possible Black Hole (BH) that can be described semi-classically in the EFT, without automatically violating the known entropy bounds in the presence of $N$ light species. In fact, this can be used to \emph{define} the number of species in the presence of towers of states, which generically appear at the boundaries of moduli space and in the context of the Distance Conjecture. This definition has been shown to agree with the naive counting of weakly coupled species as long as the towers involved are sufficiently well separated (as e.g., for KK-like towers). It is clear, though, that these two approaches are not completely independent from each other, since higher derivative corrections are known to modify the area of the horizons in BH solutions and therefore their corresponding Bekenstein-Hawking entropy \cite{Sen:2005wa}. In fact, they have been shown to be consistent via general EFT analysis in \cite{vandeHeisteeg:2022btw}, and also in the context of generation of BH horizons of species scale size in \cite{Calderon-Infante:2023uhz}.

The evidence for the species scale as a quantum gravity cut-off is moreover strongly supported in asymptotic regions of the moduli spaces pertaining to String Theory, where it seems to capture the relevant quantum gravity scale associated to the dual descriptions that arise as the infinite distance points are approached. In particular, it agrees with the higher-dimensional Planck scale in (dual) decompactification limits or rather with the string scale when probing emergent string limits \cite{Castellano:2022bvr, vandeHeisteeg:2022btw, Calderon-Infante:2023ler}, which are believed to be the only two possible ones according to the Emergent String Conjecture \cite{Lee:2019wij}.

Most of these species scale computations precisely take place close to the infinite distance boundaries of moduli space, where the weak coupling behaviour helps in determining the spectrum of the theory as well as organising the EFT expansion. However, in a recent work \cite{vandeHeisteeg:2022btw}, the authors gave a first proposal for a globally defined species scale in 4d $\mathcal{N}=2$ theories from precisely identifying the scale suppressing the first higher-curvature correction, which happens to be exactly computable in this setup. It is given by the $\mathcal{F}_1$ coefficient accompanying the quadratic curvature correction originally calculated in \cite{Gopakumar:1998ii, Gopakumar:1998jq}.

In this regard, it seems reasonable to try to push the idea of a global definition of species scale valid across the whole moduli space from studying the higher derivative corrections that are known to appear in different String Theory examples. In this context, even though the Distance Conjecture seems to be related only to infinite distance limits, where the interplay between the infinite towers of light states is the key to identify the species scale, the concept motivating such conjecture, namely dualities, can also shed some light into the problem of defining a global species scale across all moduli space. Our goal is then to use the knowledge of higher-derivative corrections, together with dualities (in particular $SL(n, \mathbb{R})$) to try to propose a well-defined global notion of species scale.

Let us mention in passing that, in principle, several different kinds of corrections can arise in a gravitational EFT, and not all of them need to be a priori suppressed by a single UV scale (as we will see in explicit examples below). Still, this should not be regarded as a drawback in the identification of the species scale from higher-curvature operators, since it is not generally expected that all of these corrections be sensitive to the full number of states in the theory (and thus to the full gravitational theory). In particular, we find that the ones that seem to be suppressed by the species scale are the UV divergent terms that must be regularized \cite{Green:1997as}, whereas some other finite corrections might be related to scales of other states, not necessarily the ones that become massless along the asymptotic limits.

\subsection{Finding a globally defined Species Scale from dualities}
\label{ss:modinvariantss}


In the remainder of this section we entertain ourselves exploring the idea of whether duality symmetries alone, together with suitable asymptotic behaviour, may be enough to completely fix the form of the species scale throughout the entire moduli space. To do so, we focus on the simpler $SL(2,\mathbb{Z})$ case and use it as a proof of concept. Let us remark that the results from this section can be thought of as complementary but independent to the rest of the paper. 
The conclusion will be that, even though it is indeed not possible to single out the relevant function, one can obtain some good candidate that reproduces all qualitative features to great accuracy.

Let us then consider the case in which some sector/branch of the moduli space $\mathcal{M}$ of a supersymmetric theory is given precisely by the coset $SL(2, \mathbb{R})/U(1)$, which we parametrise by the complex scalar $\tau=\tau_1 + \text{i} \tau_2$. The motivation for choosing such particular example comes from both its simplicity and the fact that it appears in many instances across the string Landscape. For concreteness, one can think of explicit realisations such as e.g., the \emph{full} moduli space of the 10d Type IIB String Theory or some piece of the vector multiplet moduli space of 4d $\mathcal{N}=2$ compactifications of the heterotic string on $K3\times T^2$ \cite{Kaplunovsky_1995}. We moreover assume that our theory presents some discrete $SL(2, \mathbb{Z})$ duality symmetry, which means that the true moduli space is described as the coset $\mathcal{M}= SL(2, \mathbb{Z}) \backslash SL(2, \mathbb{R})/U(1)$. Our aim is thus to propose a universal form for the species scale $\LQG (\tau)$ defined over $\mathcal{M}$.

We will restrict ourselves to the fundamental domain $\mathscr{F}$ in what follows:
\begin{align}\label{funddomain}
\mathscr{F} = \left\{\tau\in \mathbb{C} \;\;\vline \;\; |\tau|\geq 1 \, ,\; -\frac{1}{2} \leq \text{Re}\,\tau \leq \frac{1}{2}\right\}\, . 
\end{align}
Furthermore, whenever the moduli space is described as the coset $SL(2, \mathbb{R})/U(1)$ one can find a parametrisation that leads to the following kinetic term for the modulus $\tau$
\beq
\mathcal{L}_{\text{scalar}} = -\frac{\partial \tau \cdot \partial \bar \tau}{4\tau_2^2} \ .
\label{eq:scalarlag}
\eeq
Such moduli space presents three singularities within the fundamental domain: one at infinite distance ($\tau \to \text{i} \infty$) and two cusps (at $\tau=\text{i}, e^{\frac{2\pi \text{i}}{3}}$). The question is then what kind of function $\mathcal{G}(\tau, \bar \tau)$ could give rise to a properly behaved $\LQG$. In principle, it seems natural to ask for at least two things: 

\begin{enumerate}[i)]
    \item $\LQG$ must be bounded from above (since it cannot exceed $\Mpd$) and it should vanish asymptotically at infinite distance, namely $\mathcal{G} (\tau, \bar \tau) \to 0$ as $\tau \to \text{i} \infty$.

    \item It should be an \emph{automorphic} form, namely a modular invariant function of $\tau$ satisfying
\beq\label{eq:modtransf}
 \mathcal{G} \left(\frac{a \tau +b}{c\tau +d}\, , \, \frac{a \bar \tau +b}{c \bar \tau +d} \right) = \mathcal{G} (\tau, \bar \tau)\, , \qquad ad-cd=1\, ,
\eeq
where $a,b,c,d \in \mathbb{Z}$.
\end{enumerate}

At this point, one may wonder whether these two conditions are restrictive enough so as to single out the function $\mathcal{G}(\tau, \bar \tau)$ we seek for. Unfortunately, the answer turns out to be negative, although one can still extract some useful information about its possible behaviour and dependence with respect to the modulus $\tau$ by adding some additional physical requirements. For example, one could imagine such modular invariant species scale $\LQG$ to be given by certain \emph{holomorphic} function of $\tau$. However, it is easy to argue that this is actually incompatible with properties \emph{i)} and \emph{ii)} above, since there is no weight 0 (i.e. automorphic), non-singular modular form which is non-constant. This follows essentially from Liouville's theorem, and as a corollary one can argue that any meromorphic modular function of zero weight must be of the form $ \mathcal{G} \left(\tau\right) = \frac{P(j(\tau))}{Q(j(\tau))}\, $ \cite{diamond2006first}, with $P(x)$ and  $Q(x)$ some arbitrary polynomials in $j(\tau)$. Hence, from the fundamental theorem of calculus, such function $\mathcal{G}$ would necessarily have some pole(s) in the interior of $\mathscr{F}$ given precisely by the pull-back to the $\tau$-plane of the roots of the $Q$-polynomial. We thus conclude that $\mathcal{G}$ \emph{must} must be non-holomorphic in $\tau$, namely $\partial_{\bar \tau}\LQG(\tau, \bar \tau) \neq 0$, which in turn leaves room for many possible functions fulfilling the above criteria, as we explain below. 

Up to this point, we have been able to pin down the global definition of the species scale in the presence of an $SL(2,\mathbb{Z})$ to a modular invariant, non-holomorphic function of $\tau$ fulfilling conditions $i)$ and $ii)$ above. This has been achieved based purely on minimal physical requirements, such as having the correct asymptotic behaviour or the absence of singularities and non-trivial moduli dependence. This allows us to constrain its form enormously, while not being able to give a unique closed expression. For this reason, we proceed in the following to present an educated guess for the latter, based on the aforementioned physical arguments together with some already known results. In upcoming sections of this work we will confront our ansatz with various String Theory setups, thus showing complete or at least qualitative agreement with the observed behaviour therein. 

\subsubsection*{A potential candidate for $\mathcal{G}(\tau, \bar \tau)$}
\label{ss:proposalf}

We will now propose some ansatz for our modular invariant species scale in the sector parameterized by $\tau$, which takes the following form
\beq\label{eq:modspeciesscale}
\left( \frac{\LQG(\tau, \bar \tau)}{\Mpd} \right)^{d-2} = \frac{1}{\alpha \left[-\text{log} \left(\tau_2\,|\eta(\tau)|^4\right)\right]^{\gamma} + \beta}\, ,
\eeq
where $\eta(\tau)$ is Dedekind eta function (c.f. eq. \eqref{eq:Dedekind}) and $\alpha, \beta, \gamma \in \mathbb{R}_{\geq 0}$. The particular functional form chosen in eq. \eqref{eq:modspeciesscale} above is also motivated by the modular examples in 4d $\mathcal{N}=2$ models discussed later on (see eq. \eqref{4dtopologicalfreeenergy}). 

Equivalently, what we propose is the following moduli dependence for the number of light species in a theory enjoying $SL(2, \mathbb{Z})$ invariance:
\beq\label{eq:Nspecies}
N = \alpha \left[-\text{log} \left(\tau_2\,|\eta(\tau)|^4\right)\right]^{\gamma} + \beta\, ,
\eeq
which is of course a non-holomorphic function on $\tau$ as well as automorphic. It moreover grows as $\tau_2^{\gamma}$ whenever $\tau \to \text{i} \infty$ (see eq. \eqref{eq:asymptotic behavior}), which indeed tells us that $\LQG(\tau, \bar \tau) \sim \tau_2^{-\frac{\gamma}{d-2}} \to 0$ (in Planck units) upon taking $\tau_2 \to \infty$. Therefore it fulfills the two criteria discussed around eq. \eqref{eq:modtransf}, and one can check that there are no singularities within $\mathscr{F}$. The physical meaning of the free parameters $\alpha, \beta, \gamma$ is transparent: $\gamma$ controls the asymptotic decay rate for the species scale with respect to the moduli space distance (which indeed determines the relevant physics of the infinite distance limit), whilst $\beta$ and $\alpha$ are relevant for the behaviour of $\LQG(\tau, \bar \tau)$ in the interior of moduli space, determining e.g., the number of species at the desert point\cite{vandeHeisteeg:2022btw}.

\label{ss:G2}
	
Let us finally explore whether we can obtain some closed form for $\partial_{\tau} \log \LQG$ in terms of modular functions. For large modulus, the prediction is that this expression should indeed go to a constant, which corresponds to the modulus of the so-called species vector, defined in analogy to the `scalar charge-to-mass' ratio for a scalar particle \cite{Calderon-Infante:2023ler} (see also the discussion around eq. \eqref{eq:speciescalechargetomass} below), and one could thus expect to be able to constrain $\gamma$ from imposing this. We will see that, even though this does not fix $\gamma$ to a particular value, it does correlate it to the number of dimensions of the EFT and the particular infinite distance limits that are probed asymptotically.
	
For simplicity, let us take for $N$ the following expression (c.f. eq. \eqref{eq:modspeciesscale})
\beq
	 N = \left(-\log(\tau_2|\eta(\tau)|^4 \right)^\gamma\, .
\eeq
Note that adding an overall constant multiplying this expression would change nothing, since we are interested just in ratios. Analogously, an additive piece would play no role in the large modulus limit, where the $\eta$-term clearly dominates (see eq. \eqref{eq:asymptotic behavior}). Therefore, upon taking derivatives with respect to $\tau$ we obtain
\beq
	\frac {\partial N}{\partial \tau} =  \, \frac{\gamma}{2\pi \text{i}} \left( -\log(\tau_2|\eta(\tau)|^4 \right)^{\gamma -1} \,  {\tilde G}_2(\tau, \bar\tau) , 
\eeq
where we have used that $\partial_{\tau}\eta(\tau)=\left(-4\pi \text{i}\right)^{-1}\eta(\tau)G_2(\tau)$ \cite{Cvetic:1991qm}, and with ${\tilde G}_2$ being the Eisenstein (non-holomorphic) modular form of weight 2, which is defined as follows\footnote{\label{fnote:Eisenstein}The Eisenstein series of weight 2 can be computed using the sum representation
\beq
G_{2}(\tau) = 2\zeta(2) + 2\sum_{n=1}^{\infty} \sum_{m \in \mathbb{Z}} \frac{1}{\left( m+n\tau\right)^{2}}\, ,
\eeq since the usual expression \eqref{eq:holoEisenstein} is not well-defined for $k=1$ due to the lack of absolute convergence of the series. Actually, $G_2$ is not a modular form and it transforms as $G_2 \left(\frac{a \tau +b}{c\tau +d} \right)= \left( c\tau +d \right)^{2} G_{2}(\tau) -2 \pi \text{i} c \left( c\tau +d \right)$ under $SL(2, \mathbb{Z})$.} 
\beq\label{eq:nonholoG2}
	{\tilde G}_2 (\tau, \bar \tau)=G_2 (\tau)-\frac{\pi}{\text{Im}\, \tau}\, .
\eeq
Thus, we have in the end
\beq
	\frac {\partial N}{\partial \tau}\ 
 =\ \frac {\gamma}{2\pi \text{i}} N^{\frac{\gamma-1}{\gamma}}\, {\tilde G}_2\, .
\eeq
Using the definition \eqref{eq:speciesscale}, this allows us to compute  the logarithmic derivatives, which gives
\beq
	\frac {1}{\LQG} \frac {\partial \LQG}{\partial \tau} = 
 -\, \frac {\gamma }{2\pi \text{i}(d-2)}\ N^{-\frac{1}{\gamma}} \ {\tilde G}_2\, .
\eeq
Note, however, that this quantity is computed via taking some holomorphic derivative, but here we are more interested in the canonically normalised field $\hat \tau$ associated to the imaginary part of $\tau$, namely $\tau_2$. Indeed, consider a kinetic lagrangian of the form
\beq \label{eq:lagrangianII}
	\mathcal{L}_{\text{scalar}} =  -c\, \frac{\partial \tau \cdot \partial \bar \tau}{4\tau_2^2}\, ,
\eeq
where we have added some constant $c \in \mathbb{R}$ for generality. Then a canonical kinetic term for a field $\hat{\tau}$ with $\tau_2=e^{a\hat{\tau}}$ is obtained for the particular choice $a=\sqrt{2/c}$. Consequently, $\partial \hat{\tau} /\partial \tau_2 = \sqrt{c}/\tau_2 \sqrt{2}$, such that
\beq
	-\, \frac {1}{\LQG} \frac {\partial \LQG}{\partial \hat{\tau}} = \sqrt{\frac{2}{c}}\, \frac {\gamma}{\pi (d-2)} \frac {\tau_2\, \text{Re}\, {\tilde G}_2}{\left(-\log(\tau_2|\eta(\tau)|^4 \right)}\, .
\eeq
According to the Distance conjecture, for large moduli this should go to a constant, in order to properly define a convex hull \cite{Etheredge:2022opl, Calderon-Infante:2023ler}. It is then easy to see, upon using the large $\tau_2$ limit
\beq \label{eq:larget2limits}
	-\log \left(\tau_2|\eta(\tau)|^4 \right)\, \to \, \frac {\pi}{3}\tau_2\, , \qquad {\tilde G}_2\ \to \frac {\pi^2}{3}\, ,
\eeq
that a constant is obtained regardless of the particular value of the parameter $\gamma$, as previously announced:
\beq \label{eq:chargetomassratio}
	-\, \frac {1}{\LQG} \frac {\partial \LQG}{\partial \hat{\tau}}\, \to\, \sqrt{\frac{2}{c}}\, \frac {\gamma}{d-2}\, .
\eeq
Still, this quantity has been argued to take  the following form \cite{Calderon-Infante:2023ler, vandeHeisteeg:2023ubh}
\begin{equation}
\label{eq:lambdaspecies}
\partial_{\hat{\tau}} (\log \LQG) \to \lambda_{\text{sp}}=\sqrt{\frac{p}{(d+p-2)(d-2)}}\, ,  
\end{equation} 
where $p$ represents the number of (unwarped) dimensions that are decompactified along a particular infinite distance limit that is being explored, and it can also capture the emergent string case with $p\to \infty$ . Thus, $\gamma$ cannot be restricted to a particular value but it is in turn related to these parameters characterising the infinite distance singularities present in the aforementioned subsector of the moduli space. Finally, notice that the result matches the same quantity computed upon imposing the asymptotic approximation $\LQG(\tau, \bar \tau) \sim \tau_2^{-\frac{\gamma}{d-2}}$ from the beginning.

\section{Higher dimensional String Theory Examples}
\label{s:Examples}

In this section we will try to verify whether the general field theoretic expectations for a gravitational EFT discussed in Section \ref{s:speciesscale} are realised in specific String Theory constructions. The examples from Sections \ref{ss:10dIIB}\,--\,\ref{ss:MthyT3} correspond to maximally supersymmetric theories in ten, nine and eight spacetime dimensions, respectively. The latter are highly constrained, such that certain higher-dimensional (protected) operators can be computed exactly, and therefore analysed. The focus will be placed on whether these gravitational operators present a moduli dependent behaviour compatible with eq. \eqref{eq:gravEFTexpansion} above.

\subsection{ 10d Type IIB String Theory}
\label{ss:10dIIB}

As our first example, we consider Type IIB String Theory in ten dimensions. The bosonic (two-derivative) pseudo-action in the Einstein frame is of the form \cite{Polchinski:1998rr} 
\begin{equation}\label{eq:IIB10d}
			\begin{aligned}
				S_\text{IIB}^{\text{10d}}\, =\, & \frac{1}{2\kappa_{10}^2} \int d^{10}x\sqrt{-g} \left(R-\frac{1}{2}(\partial \phi)^2\right) -\frac{1}{4\kappa_{10}^2}\int e^{-\phi} H_3\wedge \star H_3 \\
				&-\frac{1}{4\kappa_{10}^2}\int \left[e^{2 \phi}F_1 \wedge \star F_1 + e^{\phi} \tilde F_3 \wedge \star \tilde F_3 + \frac{1}{2} \tilde F_5 \wedge \star \tilde F_5  +C_4\wedge H_3 \wedge F_3\right]\,,
			\end{aligned}
\end{equation}
with the different field strengths defined as follows: $H_3=dB_2$, $\tilde{F}_1=d C_0$, $\tilde{F}_3=dC_2-C_0 H_3$ and $\tilde{F}_5= dC_4-\frac{1}{2} C_2 \wedge F_3 + \frac{1}{2} B_2 \wedge H_3$. 
The above expression is famously invariant under the non-compact group $SL(2,\mathbb{R})$. However, the full \emph{quantum} theory is only expected to preserve a discrete $SL(2,\mathbb{Z})$ symmetry due to D-instanton effects. Indeed, one can rewrite the action in eq. \eqref{eq:IIB10d} in a slightly different way
\begin{equation}\label{eq:IIB10dSL2}
			\begin{aligned}
				S_\text{IIB}^{\text{10d}}\, =\, & \frac{1}{2\kappa_{10}^2} \int d^{10}x\sqrt{-g} \left(R-\frac{\partial \tau \cdot \partial \bar \tau}{2 (\text{Im}\, \tau)^2}\right) -\frac{1}{4\kappa_{10}^2}\int \frac{1}{\text{Im}\, \tau} G_3\wedge \star \bar{G}_3 + \frac{1}{2} \tilde F_5 \wedge \star \tilde F_5\\
				&+\frac{1}{8 \text{i} \kappa_{10}^2}\int \frac{1}{\text{Im}\, \tau} C_4\wedge G_3 \wedge \bar{G}_3\, ,
			\end{aligned}
\end{equation}
where $\tau=C_0 + \text{i} e^{-\phi}$ is the axio-dilaton and $G_3=dC_2 - \tau dB_2$. The modular transformations then act on the Type IIB variables as follows \cite{Schwarz:1995dk}
\begin{align}\label{eq:SdualitytransIIB}
	&\tau \rightarrow \frac{a\, \tau + b}{c\, \tau+d}\,,\qquad \begin{pmatrix}
		C_2\\ B_2
	\end{pmatrix}
	\rightarrow \mathcal{A} \begin{pmatrix}
		C_2\\ B_2
	\end{pmatrix}\, , \qquad \mathcal{A}= \begin{pmatrix}
		a \quad  b\\c \quad  d
	\end{pmatrix} \in SL(2,\mathbb{Z})\, , \notag\\
 &C_4 \rightarrow C_4\, , \qquad g_{\mu \nu} \rightarrow g_{\mu \nu}\, ,
\end{align}
which, as can be easily checked, leave the action \eqref{eq:IIB10dSL2} invariant.

Notice that there is only one dimensionful quantity entering into the supergravity action, namely the Planck mass. This is imposed by having maximal supersymmetry, thus preventing us from obtaining further information about the quantum gravity cut-off. Following our general discussion from Section \ref{s:speciesscale}, what we should do is look at higher-dimensional operators involving the Riemann tensor, which are expected to be suppressed by some UV scale, i.e. the species scale.

Before doing so, let us take advantage of our analysis from Section \ref{ss:modinvariantss} to get a feeling of the kind of modular function that should control the behaviour of the species scale in the present setup. Recall that such function must respect the duality symmetries of the theory, and thus it should be given by some sort of automorphic form. A potential candidate for $\LQG$ would be the following
\beq\label{eq:modspeciesscaleIIB}
\left( \frac{\LQG(\tau, \bar \tau)}{M_{\text{pl},\, 10}} \right)^{8} = \left(-\text{log} \left(\tau_2\,|\eta(\tau)|^4\right)\right)^{-\gamma}\, , \qquad \text{with}\ \ \gamma > 0\, ,
\eeq
up to finite constants that can be ignored for our purposes here (namely $\alpha$ and $\beta$ in eq. \eqref{eq:modspeciesscale}). In fact, one can easily argue that for the present setup, $\gamma$ should be equal to 2. To see this, we compute the asymptotic decay rate (i.e. at infinite distance), $\lambda_{\text{sp}}$, for the species scale thus defined:
\beq
\lambda_{\text{sp}} = \left| \frac{\sqrt{G^{\tau_2 \tau_2}} \partial_{\tau_2} \LQG}{\LQG}\right|\, ,
\eeq
where $G_{\tau_2 \tau_2} = \frac{1}{2\tau_2^2}$ is the moduli space metric associated to the imaginary part of the modulus $\tau$ and $G^{\tau_2 \tau_2}$ denotes its inverse.
Notice that we are ignoring here the axionic contribution to the decay rate $\lambda_{\text{sp}}$ since it goes to zero as we approach the infinite distance boundary \cite{Calderon-Infante:2023ler}. Now, for the case at hand the singularity corresponds to an emergent string limit, so following the reasoning around eq. \eqref{eq:chargetomassratio}, we find $\lambda_{\text{sp}}=1/\sqrt{8}$ \cite{Etheredge:2022opl}, which implies that $\gamma=2$. 

Note that the species scale function such defined not only behaves asymptotically as one would expect, but it also provides an absolute maximum value for $\LQG$ (precisely when $\tau= e^{\frac{2\pi \text{i}}{3}}$) compatible with the analysis of \cite{Long:2021jlv} (see also \cite{vandeHeisteeg:2022btw}), where it was argued that the `desert point' should happen when the BPS gap of $(p,q)$-strings is maximized.

\subsubsection*{The $R^4$ operator}

In order to check how these EFT expectations are furnished in Type IIB String Theory, we will look at the first non-trivial higher-curvature operator appearing in the effective action. Such correction is BPS-protected --- so we can be sure that it receives no further quantum contributions, involves four powers of the Riemann tensor and has the form (in the Einstein frame) \cite{Green:1999pv,Green:1997di,Pioline:1998mn}
\beq
S_{R^4}^{\text{10d}}= \frac{1}{\ell_{10}^2} \int d^{10}x \sqrt{-g}\, E_{3/2}^{sl_2} (\tau, \bar \tau)\, t_8 t_8 R^4\, ,
\label{eq:10dR^4IIB}
\eeq
where $t_8 t_8 R^4 \equiv t^{\mu_1 \ldots \mu_8} t_{\nu_1 \ldots \nu_8} R^{\nu_1 \nu_2}_{\mu_1 \mu_2} \ldots R^{\nu_7 \nu_8}_{\mu_7 \mu_8}$, with the tensor $t^{\mu_1 \ldots \mu_8}$ defined in \cite{Green:1981ya} (see also Appendix 9 of \cite{Green:2012pqa}).\footnote{Notice that the quantity $t_8 t_8 R^4$ transforms as $t_8 t_8 R^4 \to e^{-2\phi} t_8 t_8 R^4$ upon performing the Weyl rescaling to the 10d string frame metric $g_{\mu \nu} \to e^{\phi/2} g_{\mu \nu}$.} Here $E_{3/2}^{sl_2} (\tau, \bar \tau)$ denotes the order-$3/2$ non-holomorphic Eisenstein series of $SL(2,\mathbb{Z})$, which is an automorphic function that can be defined in a series expansion of the complex valued field $\tau$ as follows (see Appendix \ref{ap:Massform} for details)
\begin{align}\label{eq:nonpertexpansionE3/2}
	E_{3/2}^{sl_2} =\, 2\zeta(3) \tau_2^{3/2} + 4\zeta(2) \tau_2^{-1/2} + \mathcal{O} \left( e^{-2\pi \tau_2}\right)\, .
\end{align}
Due to automorphicity, it is enough to restrict ourselves to the fundamental domain $\mathscr{F}$ of $SL(2,\mathbb{Z})$ (c.f. eq. \eqref{funddomain}) when studying the asymptotic behaviour of $E_{3/2}^{sl_2}$. This leaves us with only one possible infinite distance limit, namely the weak coupling point $\tau_2 \to \infty$, such that, at leading order, the $R^4$-correction behaves like $\tau_2^{3/2}$ for large $\tau_2$.

How does this fit with our general considerations from Section \ref{s:speciesscale}? Notice that the $R^4$-term has mass dimension $n=8$, such that upon comparing with the expected behaviour for higher-dimensional operators in a gravitational EFT in $d$ spacetime dimensions with a schematic lagrangian of the form (see discussion around eq. \eqref{eq:gravEFTexpansion})
\beq
\mathcal{L}_{\mathrm{EFT,\,}d} \supset \frac{1}{2 \kappa_d^2} \left[ \left(R + \sum_n \frac{\mathcal{O}_n (R)}{\LQG^{n-2}}\right) -\frac{\partial \tau \cdot \partial \bar \tau}{2\tau_2^2}\right]\, ,
\label{eq:scalargravDlag}
\eeq
we conclude that the coefficient accompanying such term in the effective action should grow like $\LQG^{-6}\, R^4$ in Planck units. The species scale in the present case is identified with the string scale, which in Planck units is given by
\beq\label{eq:stringscale}
m_s=\frac{M_{\text{pl,\,} 10}}{\left(4\pi \tau_2^{2}\right)^{1/8}} \, ,
\eeq
such that $\LQG^{-6} \sim M_{\text{pl,\,} 10}^{-6}\, \tau_2^{3/2}$, in agreement with eq. \eqref{eq:nonpertexpansionE3/2}.

\subsubsection*{Further tests}

One can try to go further with this analysis by looking at the next few known contributions to the four-(super)graviton effective action in $\mathcal{N}=(0,2)$ 10d maximal supergravity. These terms --- which still preserve some reduced amount of supersymmetry --- involve respectively four and six derivatives of $R^4$ and they receive both perturbative and non-pertutbative corrections. The first of them, which corresponds to a gravitational operator of mass dimension $n=12$, reads as \cite{Green:1999pu}
\beq
S_{\partial^4R^4}^{\text{10d}} = \frac{\ell_{10}^2}{2} \int d^{10}x \sqrt{-g}\,E_{5/2}^{sl_2} (\tau, \bar \tau)\, \partial^4 R^4\, ,
\label{eq:10dpartial4R^4IIA}
\eeq
whose moduli dependence arises from the order-$5/2$ non-holomorphic Eisenstein series, thus being compatible with the $SL(2, \mathbb{Z})$ invariance of the theory.  As it was also the case for the $R^4$-term before, in order to check whether the expected expansion \eqref{eq:scalargravDlag} also holds for this case we only need to look at the large $\tau_2$ behaviour. Upon doing so, one finds (c.f. eq. \eqref{eq:nonpertexpansion})
\beq
\mathcal{L}_{\partial^4R^4}^{\text{10d}} = \frac{\ell_{10}^2}{2} \left( 2\zeta(5) \tau_2^{5/2} + \frac{4\pi^4}{135} \tau_2^{-3/2} + \mathcal{O}(e^{-4\pi\tau_2}) \right) \partial^4 R^4\, ,
\eeq
which to leading order agrees with $\LQG^{-10}\, \partial^4 R^4$, where $\LQG=m_s$ (c.f. eq. \eqref{eq:stringscale}).

On the other hand, the second term involving six derivatives of the Riemann tensor reads as follows\cite{Green:2005ba,Green:2010wi,Green:2010kv}
\beq
S_{\partial^6R^4}^{\text{10d}} = \ell_{10}^4 \int d^{10}x \sqrt{-g}\, \mathcal{E} (\tau, \bar \tau)\, \partial^6 R^4\, .
\label{eq:10dpartial6R^4IIA}
\eeq
where $\mathcal{E} (\tau, \bar \tau)$ is a modular form which does not belong to the non-holomorphic Eisenstein series. It can be nevertheless expanded around $\tau_2\to \infty$, yielding
\beq
\mathcal{L}_{\partial^6R^4}^{\text{10d}} = \ell_{10}^4 \left( \frac{2\zeta(3)^2}{3} \tau_2^{3} + \frac{4 \zeta(2) \zeta(3)}{3}\tau_2 + \frac{8\zeta(2)^2}{5}\tau_2^{-1} + \frac{4\zeta(6)}{27}\tau_2^{-3} + \mathcal{O}(e^{-4\pi\tau_2})\right) \partial^6 R^4\, ,
\eeq
where the first term corresponds to the tree-level contribution, whilst the remaining pieces --- except for the exponentially suppressed correction --- include up to three-loop contributions in $g_s$ (see \cite{Green:2010wi} and references therein). Notice that since the above operator has mass dimension $n=14$, one expects according to eq. \eqref{eq:scalargravDlag} a dependence of the form $\LQG^{-12}\, \partial^6 R^4$, which indeed matches asymptotically with the species scale computation and the EFT expectations (see eq. \eqref{eq:scalargravDlag} above).

Beyond four-point graviton scattering one may also consider higher-dimensional operators mixing both the gravitational field and the Ramond-Ramond $p$-forms. In particular, there is an infinite family of such terms in the 10d Type IIB effective action, which are of the form $R^4 |G_3|^{4g-4}$. These, upon compactification on any hyper-K\"ahler manifold, can be alternatively computed in terms of the $\mathcal{N}=4$ topological String Theory \cite{Berkovits:1994vy} (see also \cite{Ooguri:1995cp}).\footnote{A similar phenomenon happens in 4d $\mathcal{N}=2$ theories, where certain higher derivative F-terms can be computed by the $\mathcal{N}=2$ topological string, see Section \ref{s:4dN=2} below.} Moreover, their precise moduli dependence has been conjectured to be of the form \cite{Berkovits:1998ex}
\begin{equation}\label{eq:BerkovitsVafa}
			\begin{aligned}
				S_\text{IIB}^{\text{10d}}\, \supset\, & \int d^{10}x\sqrt{-g}\, R^4 \sum_{g\geq1} \ell_{10}^{4g-6} \alpha_{g} \sum_{k=2-2g}^{2g-2} (-1)^k \tau_2^{-2g+2}G_3^{2g-2+k} \bar{G}_3^{2g-2-k}\\
                &\sum_{(m, n) \in \mathbb{Z}^2 \setminus \lbrace (0,0) \rbrace} \frac{\tau_2^{g+1/2}}{\left( m+n\tau\right)^{g+1/2+k} \left( m+n \bar \tau\right)^{g+1/2-k}}\, ,
			\end{aligned}
\end{equation}
where $\alpha_g$ denote some unknown normalization coefficients. Notice that for $k=0$, the operators are simply $R^4 \left|\tau_2^{-1/2} G_3 \right|^{4g-4}$. They are manifestly modular invariant, have mass dimension $n=4g+4$ and their corresponding Wilson coefficients become $E_{g + 1/2}^{sl_2} (\tau, \bar \tau)$. Therefore, at infinite distance each of these higher-derivative terms behave like $\tau_2^{g +1/2} \sim m_s^{-4g-2}$ (in 10d Planck units), in perfect agreement with \eqref{eq:scalargravDlag}.

\subsubsection*{A Closer Look into the EFT Expansion}

There are a couple of lessons that one can extract from the previous Type IIB example. First, notice that our candidate \eqref{eq:modspeciesscaleIIB} does not match with the exact result unless we take the infinite distance limit $\tau_2 \to \infty$. This may suggest that, in general, our ansatz cannot give the correct answer, although it may serve as a good proxy to characterise the qualitative behaviour of the species scale, even within the bulk of the moduli space. Instead, what one should take here as the species scale globally defined would be the following modular expression
\beq\label{eq:speciesIIB}
\LQG = \left(E_{3/2}^{sl_2} (\tau, \bar \tau)\right)^{-1/6}\, ,
\eeq
which also satisfies the two minimal requirements for defining a bona-fide species scale, see the discussion around eq. \eqref{eq:modtransf}. 

Another interesting fact about the higher-curvature corrections here described is that they do not strictly organise in powers of the cut-off function \eqref{eq:speciesIIB}, as the naive expectation from eq. \eqref{eq:scalargravDlag} suggests. Indeed, such moduli-dependent functions are seen to be given by certain automorphic forms of $SL(2,\mathbb{Z})$ (satisfying some eigenvalue equation \cite{Green_1999}), which cannot be written in general as powers of one another. This means that perhaps the sensible way to think about the gravitational EFT expansion is in terms of some `perturbative' approximation, which may hold only \emph{asymptotically} in moduli space, where one usually recovers the weak coupling behaviour. In fact, it is precisely close to the infinite distance point where one sees the cut-off expansion in a power series to emerge, with further suppressed perturbative and non-perturbative quantum corrections (in the appropriate dual frame), which can be interpreted as some sort of `anomalous dimensions'.

\subsection{ 10d Type IIA String Theory}
\label{ss:10dIIA}

We turn now to ten-dimensional Type IIA String Theory, whose bosonic action is given in the Einstein frame by \cite{Polchinski:1998rr}
\begin{equation}\label{eq:IIA10d}
			\begin{aligned}
				S_\text{IIA}^{\text{10d}} =\frac{1}{2\kappa_{10}^2} \int d&^{10}x\sqrt{-g} \left(R-\frac{1}{2}(\partial \phi)^2\right)-\frac{1}{4\kappa_{10}^2}\int e^{-\phi} H_3\wedge \star H_3 \\
				&-\frac{1}{4\kappa_{10}^2}\int \left[e^{\frac{3}{2}\phi}F_2 \wedge \star F_2 + e^{\frac{1}{2}\phi} \tilde F_4 \wedge \star \tilde F_4 + B_2\wedge F_4 \wedge F_4\right]\, , 
			\end{aligned}
\end{equation}
where $H_3=dB_2$, $F_2=d C_1$ and $\tilde{F}_4=d C_3-C_1 \wedge H_3$. This theory has a trivial U-duality group \cite{Hull:1994ys}, with a moduli space $\mathcal{M}_{\text{IIA}} \cong \mathbb{R}$ parametrised by the dilaton modulus $\phi$. Given that the theory does not enjoy $SL(2, \mathbb{Z})$ invariance, we cannot use our ansatz \eqref{eq:modspeciesscale} to predict the functional form of the species scale. However, according to our general discussion in Section \ref{s:speciesscale}, it should still be true that such quantity controls the gravitational EFT expansion. Thus, proceeding analogously as we did for Type IIB String Theory, let us look at the first non-zero correction of the previous two-derivative action. In the Einstein frame, it reads as follows \cite{Grisaru:1986dk,Grisaru:1986kw,Gross:1986iv} 
\beq
S_{R^4}^{\text{10d}}= \frac{1}{\ell_{10}^2} \int d^{10}x \sqrt{-g}\, \left( 2\zeta(3) e^{-3\phi/2} + \frac{2\pi^2}{3} e^{\phi/2}\right) t_8 t_8 R^4\, ,
\label{eq:10dR^4IIA}
\eeq
which is nothing but the expression \eqref{eq:10dR^4IIB} with the instanton sum excluded. In fact, the first term corresponds to the tree-level contribution (which arises at fourth-loop order in the 2d $\sigma$-model perturbation theory), whilst the second piece is a one-loop string correction in $g_s$. 

Let us now check what are the relevant asymptotics of this dimension-8 operator. At weak coupling, namely when $\phi \to -\infty$ (equivalently $g_s \to 0$), the tree-level term dominates, and we obtain
\beq
\mathcal{L}_{R^4}^{\text{10d}}\, =\,  \frac{2\zeta(3)}{\ell_{10}^2} e^{-3\phi/2}\, t_8 t_8 R^4 + \ldots\ .
\label{eq:R^4weakcoupling}
\eeq
Comparing this with eq. \eqref{eq:gravEFTexpansion}, we deduce that the coefficient accompanying such term in the effective action should behave as $\LQG^{-6}\, R^4$ asymptotically. Therefore, since the species scale coincides with the string scale along the weak coupling limit, we again find that
\beq
\left(\frac{\LQG}{M_{\text{pl, 10}}}\right)^{-6}\, \sim\, e^{-3\phi/2}\, ,
\eeq
in agreement with eq. \eqref{eq:R^4weakcoupling} above.

Contrary, at strong coupling, it is the one-loop correction which becomes more important, thus leading to the following dilaton dependence
\beq
\mathcal{L}_{R^4}^{\text{10d}}\, =\,  \frac{4\zeta(2)}{\ell_{10}^2} e^{\phi/2}\, t_8 t_8 R^4 + \ldots\ ,
\label{eq:R^4strongcoupling}
\eeq
whilst the species counting is now dominated by the tower of D0-brane bound states instead, since the fundamental string becomes infinitely heavy in 10d Planck units. Following the definition in \eqref{eq:speciesscale}, one recovers that the species scale is capturing the 11d M-theory Planck scale, given by \cite{Castellano:2022bvr}
\beq \label{eq:QGscaleandN} 
		\Lambda_{\text{QG}}\, \sim\, m_{\text{D}0}^{1/9}\ \Mpt^{8/9}\, \sim\,  e^{-\phi/12} \Mpt\, ,
\eeq
such that the quantity $\LQG^{-6}\, R^4$ precisely reproduces the parametric dependence in eq. \eqref{eq:R^4strongcoupling} above.

Let us also stress that the fact that Type IIA String Theory does not enjoy S-duality invariance is responsible for the two singular limits, namely that of weak/strong coupling, to behave very differently from one another (c.f. eqs. \eqref{eq:R^4weakcoupling} and \eqref{eq:R^4strongcoupling}). On the other hand, in the Type IIB setup, $SL(2, \mathbb{Z})$ invariance forces all the terms to be self-dual, which implies that the physics at weak/strong coupling are exactly the same, and the 11d Planck scale never dominates the EFT expansion. 

Finally, let us comment that by performing a similar analysis to the one done for the Type IIB case regarding the corrections of the form $\partial^4 R^4$ and $\partial^6 R^4$, it can be seen that they are not suppressed by the species scale to the expected power when the M-theory limit is probed. In contrast, different combinations involving the species scals and the mass of the D0-branes appear. The emergent string limits, however, are always suppressed by the right power of the $\LQG$, but one must take into account that in such cases the species scale and that of the tower coincide (i.e. $\LQG = m_s$). It seems reasonable to single out the $R^4$ as the one probing the QG cut-off (since it has $n<d=10$), whereas different scales can suppress other operators with higher mass dimension. We will comment further on this in Section \ref{ss:gravEFTexpansion} below.

\subsection{M-theory on $T^2$}
\label{ss:MthyT2}

Let us now turn to the unique 9d $\mathcal{N}=2$ supergravity theory, which may be obtained by e.g., compactifying M-theory on a $T^2$, whose metric can be parametrised by
\begin{equation}\label{eq:T2metric}
	g_{m n}= \frac{\mathcal{V}_2}{\tau_2} \left(
	\begin{array}{cc}
		1 & \tau_1  \\
		\tau_1 & |\tau|^2  \\
	\end{array}
	\right) \, ,
\end{equation}
with $\tau=\tau_1+{\rm i}\tau_2$ denoting the complex structure of the $T^2$ and $\mathcal{V}_2$ its overall volume (in 11d Planck units). The scalar and gravitational sectors in the 9d Einstein frame read (see Appendix \ref{ap:maxSUGRA8d} for details)
\begin{equation}\label{eq:9d}
	S_\text{M-th}^{\text{9d}} \supset \frac{1}{2\kappa_9^2} \int d^{9}x\, \sqrt{-g}\,  \left[ R - \frac{9}{14} \frac{\left( \partial \mathcal{V}_2 \right)^2}{\mathcal{V}_2^2} -\frac{\partial \tau \cdot \partial \bar \tau}{2 \tau_2^2} \right]\, .
\end{equation}
This theory has a moduli space which is classically exact and parametrises the manifold $\mathcal{M}_{\text{9d}}=SL(2, \mathbb{Z})\backslash SL(2, \mathbb{R})/U(1) \times \mathbb{R}_{+}$, where we have taken into account the $SL(2, \mathbb{Z})$ U-duality symmetry associated to the full quantum theory \cite{Schwarz:1995dk,Aspinwall:1995fw}. 

\begin{figure}[htb]
\begin{center}
\includegraphics[width=0.6\textwidth]{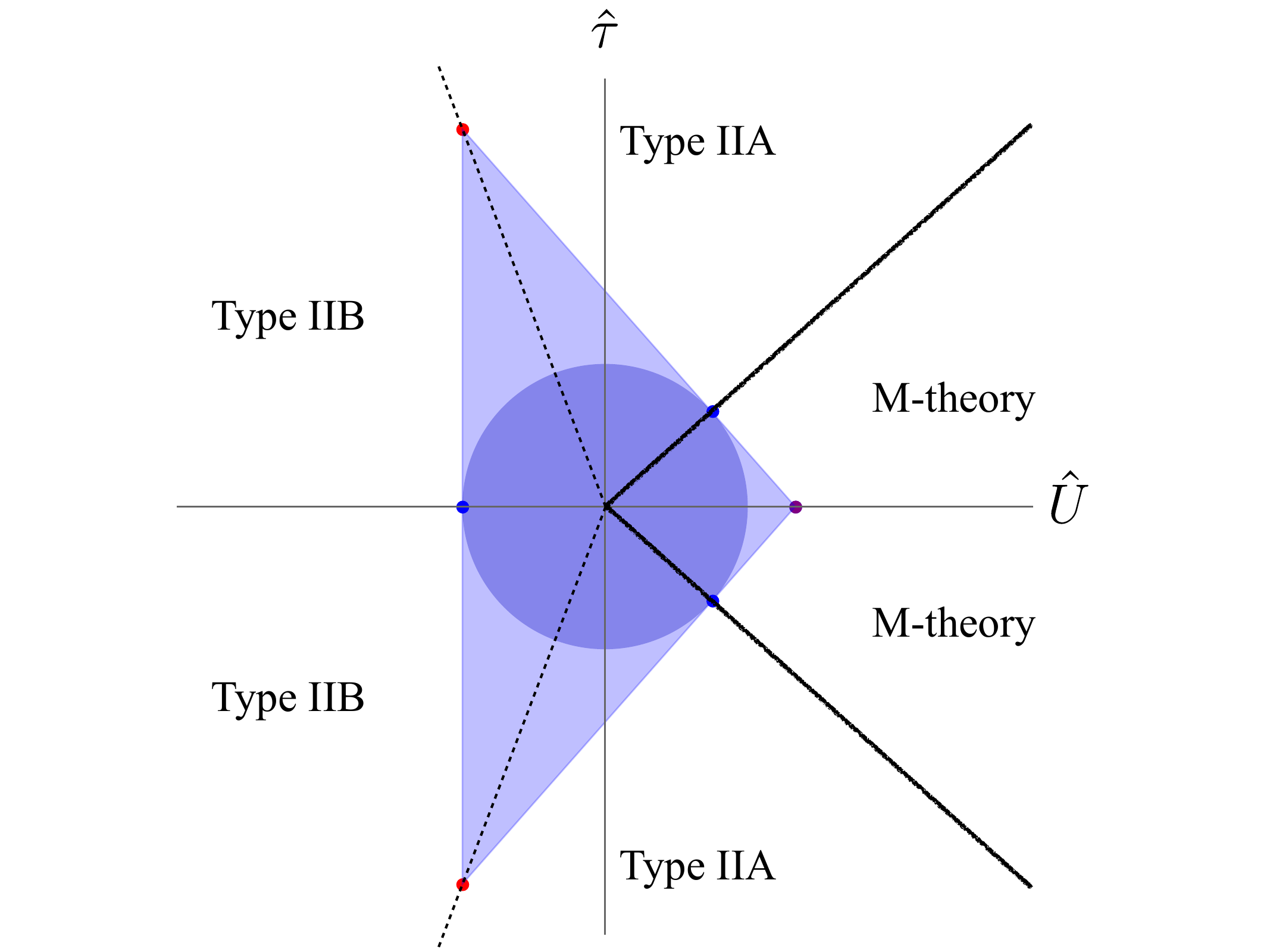}
\caption{\small Convex hull diagram for the species scale in M-theory on $T^2$ in the plane $(\hat U, \hat \tau) = (\frac{3}{\sqrt{14}} \log \mathcal{V}_2\, ,\, \frac{1}{\sqrt{2}}\log \tau_2)$. The blue dots in the facets represent the single KK towers, whereas the red and purple dots at the vertices represent string oscillator and double KK towers. The self-dual line $\hat{\tau}=0$ is fixed under the $\mathbb{Z}_2$ remnant symmetry.}
\label{fig:MthyT2}
\end{center}
\end{figure}

The key aspects regarding the (asymptotic) behaviour of the species scale function over $\mathcal{M}_{\text{9d}}$ are captured by the so-called species vectors $\vec{\mathcal{Z}}$, which determine the rate at which the species scale goes to zero asymptotically at each infinite distance boundary and are defined as follows \cite{Calderon-Infante:2023ler}
\begin{equation}\label{eq:speciescalechargetomass}
	\mathcal{Z}^a_{\text{sp}} \equiv -\delta^{ab}e_b^i\ \partial_i \log \LQG \, ,
\end{equation}
where $e_b^i$ denotes the vielbein associated to the moduli space metric $G_{ij}$. Equivalently, one can define such species vectors by taking the logarithmic derivative of $\LQG$ already in the canonically normalized frame. In fact, the present 9d example was studied in detail in \cite{Calderon-Infante:2023ler}, and the relevant species vectors are shown in Figure \ref{fig:MthyT2}. Hence, in this case, whatever the function $\LQG (\mathcal{V}_2, \tau)$ may be, it should not only go to zero asymptotically as stressed in the point $\emph{i)}$ above, but it must do so in a way that recovers the results from \cite{Calderon-Infante:2023ler}.

In order to check this, we proceed as in the 10d examples from Sections \ref{ss:10dIIB} and \ref{ss:10dIIA}. Hence, we look at the next non-trivial correction to the two-derivative lagrangian of maximal supergravity in 9d. Such operator happens to be precisely the one behaving schematically like $R^4$, and it is still BPS protected. Its dependence with respect to the moduli space parametrised by $\lbrace \mathcal{V}_2, \tau \rbrace$ has been already computed in \cite{Green:1997tv,Green:1997as}, and it is captured by the following non-trivial function (see also \cite{Green:2010wi})
\beq
S_{R^4}^{\text{9d}}= \frac{1}{\ell_{9}} \int d^{9}x \sqrt{-g}\, \left( \frac{2\pi^2}{3} \mathcal{V}_2^{6/7} + \mathcal{V}_2^{-9/14} E_{3/2}^{sl_2} (\tau, \bar \tau)\right) t_8 t_8 R^4\, ,
\label{eq:9dR^4MthT2}
\eeq
where $SL(2,\mathbb{Z})$ invariance in ensured by the non-holomorphic Eisenstein series of order $3/2$ (recall that the volume modulus is left unchanged under a modular transformation). The above operator has mass dimension $n=8$, such that according to eq. \eqref{eq:gravEFTexpansion}, we expect it to behave as $R^4 \LQG^{-6}$. We consider each asymptotic limit in Figure \ref{fig:MthyT2} in turn.

First, thanks to $SL(2, \mathbb{Z})$ invariance we can just focus on infinite distance limits where $\tau_2 \geq 0$ (i.e. those living in the upper half-plane of Figure \ref{fig:MthyT2}). Moreover, notice that as long as $\tau_2 > \mathcal{V}_2$ the second term in eq. \eqref{eq:9dR^4MthT2} dominates over the first one, which allows us to divide the `fundamental domain' in two different subregions, depending on whether $\tau_2 > \mathcal{V}_2$ or $\tau_2 < \mathcal{V}_2$.

\newpage

\underline{\textit{The M-theory regime}}

Let us start with the region $\mathcal{V}_2 > \tau_2$. In this case, the dominant term in the expression for the $R^4$-operator depends solely on the internal volume, so that one can restrict in practice to the large radius limit at fixed (and finite) complex structure. This is nothing but a full decompactification to 11d supergravity, and thus the species scale should coincide (up to order one factors) with the 11d Planck scale, which depends on the moduli fields as follows
\beq
\frac{M_{\text{pl, 11}}}{M_{\text{pl, 9}}} \sim \mathcal{V}_2^{-1/7}\, .
\label{eq:11dPlanckmass}
\eeq
Therefore, according to \eqref{eq:gravEFTexpansion} we expect the asymptotic behavior $\LQG^{-6} \sim \mathcal{V}_2^{6/7}$ in front of the quartic correction in the 9d action, which indeed matches the correct result. 

\underline{\textit{The Type II regime}}

In the opposite regime, namely when $\tau_2 > \mathcal{V}_2$, the species scale should be controlled by the fundamental string mass (defined as $m_{\text{str}} \equiv \sqrt{T_{\text{str}}}$), which corresponds to the red dot in the upper half-plane in Figure \ref{fig:MthyT2}. This can be readily computed, yielding
\beq
\frac{m_{\text{str}}}{M_{\text{pl, 9}}} \sim \mathcal{V}_2^{3/28} \, \tau_2^{-1/4}\, .
\label{eq:fundstringmass}
\eeq
Hence, focusing in the second term in \eqref{eq:9dR^4MthT2} and using the asymptotic behaviour exhibited by the order-$3/2$ non-holomorphic Eisenstein series (c.f. eq. \eqref{eq:nonpertexpansionE3/2}), we conclude that the coefficient to the $R^4$-term behaves asymptotically as $\mathcal{V}_2^{-9/14}\, \tau_2^{3/2} \sim m_{\text{str}}^{-6}$, in agreement with our prediction.

\underline{\textit{Decompactification to 10d}}

For completeness, let us also discuss the two boundaries between the different asymptotic regions in moduli space, as seen from the diagram in Figure \ref{fig:MthyT2}. These are moreover associated to certain directions signalling towards partial decompactification to either 10d Type IIA or Type IIB supergravity. On the one hand, precisely when $\tau_2 = \mathcal{V}_2 \to \infty$, a subset of KK modes become light and the theory decompactifies to 10d Type IIA supergravity. The 10d Planck scale presents the following moduli dependence
\beq
\frac{M^{\rm IIA}_{\text{pl, 10}}}{M_{\text{pl, 9}}} \sim \mathcal{V}_2^{-9/112} \tau_2^{-1/16} \sim \mathcal{V}_2^{-1/7}\, ,
\label{eq:10dPlanckmassIIA}
\eeq
which agrees asymptotically with both $M_{\text{pl, 11}}$ and $m_{\text{str}}$ along the aforementioned limit. Therefore, upon inserting $\LQG \sim M^{\rm IIA}_{\text{pl, 10}}$ into eq. \eqref{eq:gravEFTexpansion}, one obtains perfect agreement with the behaviour exhibited by \eqref{eq:9dR^4MthT2}.

On the other hand, for the limit $\mathcal{V}_2 \to 0$ the species counting is dominated by M2-branes wrapping the internal space. These states correspond to the KK tower implementing the M-theory/F-theory duality \cite{Vafa:1996xn}, such that the QG scale becomes identical to the 10d Type IIB Planck mass, which reads
\beq
\frac{M^{\rm IIB}_{\text{pl, 10}}}{M_{\text{pl, 9}}} \sim\mathcal{V}_2^{3/28}\, .
\label{eq:10dPlanckmass}
\eeq
Hence, along such limit the $R^4$-operator should be controlled by $\LQG^{-6} \sim \mathcal{V}_2^{-9/14}$, thus matching the behaviour observed in the second term in eq. \eqref{eq:9dR^4MthT2}.

Before turning to higher-dimensional operators other than $t_8 t_8 R^4$, let us make a couple of comments. Indeed, from the discussion above one would be tempted to propose $\LQG$ to be defined in the present nine-dimensional setup precisely by
\beq
\LQG = \left( \frac{2\pi^2}{3} \mathcal{V}_2^{6/7} + \mathcal{V}_2^{-9/14} E_{3/2}^{sl_2} (\tau)\right)^{-1/6}\, ,
\label{eq:9dfullspeciesfn}
\eeq
which is automorphic and moreover reproduces the correct asymptotic behaviour in every infinite distance corner of the moduli space, as we just demonstrated. However, let us stress one more time that this would imply, via eq. \eqref{eq:scalargravDlag}, that all the following higher order gravitational operators in the 9d effective action should also be accompanied by appropriate powers of the function \eqref{eq:9dfullspeciesfn}, which is known \emph{not} to be the case for e.g., higher-curvature corrections involving derivatives of the quartic curvature invariants (see below). What seems to be true, though, is the fact that the species scale, as one typically defines it close to asymptotic boundaries in moduli space (see eq. \eqref{eq:speciesscale}), is the energy scale organising the EFT expansion of our quantum-gravitational theories, and thus should be taken as the true QG cut-off.  

Second, notice that the function controlling the $R^4$-interaction here is again an eigenfunction (with eigenvalue equal to $6/7$) of the Laplace operator in $SL(2,\mathbb{R})/SO(2) \times \mathbb{R}_+$
\beq\label{eq:laplacian9d}
\Delta_{\text{9d}} = \Delta_2 - \frac{7}{9}\, \mathcal{V}_2\, \partial_{\mathcal{V}_2} \left( \mathcal{V}_2\, \partial_{\mathcal{V}_2} \right) - \frac{1}{3}\, \mathcal{V}_2\, \partial_{\mathcal{V}_2}\, ,
\eeq
with $ \Delta_2 = \tau_2^2 \left( \partial^2_{\tau_1} + \partial^2_{\tau_2}\right)$. This was also the case in the 10d Type IIB case discussed before, and the eigenvalue seems to be related to the power of the number of species that accompanies the operator under consideration.
\subsubsection*{Further tests}

Proceeding as in the previous example, let us now look at the next few contributions to the four-(super)graviton effective action in 9d $\mathcal{N}=2$ supergravity. For the four-derivative term we have \cite{Green:2010wi}
\beq
S_{\partial^4R^4}^{\text{9d}} = \ell_{9}^3 \int d^{9}x \sqrt{-g}\, \left( \frac{1}{2} \mathcal{V}_2^{-15/14} E_{5/2}^{sl_2} (\tau) + \frac{2\zeta(2)}{15} \mathcal{V}_2^{27/14} E_{3/2}^{sl_2} (\tau) + \frac{4\zeta(2)\zeta(3)}{15} \mathcal{V}_2^{-18/7}\right) \partial^4 R^4\, .
\label{eq:9dpartial4R^4}
\eeq
Notice that such term is also an eigenfunction of the scalar laplacian \eqref{eq:laplacian9d} with eigenvalue 30/7, as one can readily check. Moreover, the correction is compatible with the $SL(2, \mathbb{Z})$ invariance of the theory, such that we will restrict ourselves to the fundamental domain in what follows. The mass dimension of the operator is $n=12$, and thus, according to eq. \eqref{eq:gravEFTexpansion}, we expect an asymptotic dependence of the form $\LQG^{-10}\, \partial^4 R^4$ for \eqref{eq:9dpartial4R^4} above. In particular, this prediction is fulfilled in all the cases in which we are probing an emergent string limit, namely when $\LQG = m_{\rm str}$, whereas in decompactification limits one cannot directly identify $\LQG^{-10}$ as the coefficient in front of this correction. Still, as already mentioned, this might be understood in terms of corrections which are not only sensitive to purely quantum-gravitational effects associated with infinite towers of states becoming light asymptotically, as explained in Section \ref{ss:gravEFTexpansion} below. 

\subsection{M-theory on $T^3$}
\label{ss:MthyT3}

As our final example in this section, we consider maximal supergravity in eight spacetime dimensions. This theory arises upon compactifying e.g., Type IIB supergravity on a two-dimensional torus. Such process yields the following 8d action in the scalar-tensor sector (see Appendix \ref{ap:maxSUGRA8d} for details)
\begin{equation}\label{eq:IIB8d}
			\begin{aligned}
				S_\text{IIB}^{\text{8d}}\, \supset\, & \frac{1}{2\kappa_{8}^2} \int d^{8}x\sqrt{-g} \left[R-\frac{1}{6}\frac{(\partial \nu)^2}{\nu^2} -\frac{\partial \tau \cdot \partial \bar \tau}{2 \tau_2^2} -\frac{\partial U \cdot \partial \bar U}{2 U_2^2} - \nu \frac{\left| \tau \partial b + \partial c\right|^2}{2\tau_2}\right]\, ,
			\end{aligned}
\end{equation}
where $U$ denotes the complex structure of the internal torus, $\nu= \left( \tau_2 \mathcal{V}_2^2\right)^{-1}$ is an $SL(2, \mathbb{Z})_{\tau}\,$-invariant volume, and $b,c$ are compact scalar fields arising from the reduction of the NS and RR 2-form fields of 10d $\mathcal{N}=(2,0)$ supergravity on the internal 2-cycle. Note that there are two modular symmetries visible from the action \eqref{eq:IIB8d} above: that associated to the axio-dilaton --- which is inherited from ten dimensions, as well as an additional one which transforms the complex $U$ field in a fractional linear fashion. There is, however, an extra `hidden' $SL(2, \mathbb{Z})_{T}$ symmetry associated to the K\"ahler modulus $T=b+i\mathcal{V}_2$,\footnote{Note that upon performing a T-duality on any of the two 1-cycles within the $T^2$, one arrives at Type IIA String Theory on a (dual) torus, which is moreover described by the same exact action as in eq. \eqref{eq:IIB8d} with the $U$ and $T$ fields exchanged.} which can be made manifest upon changing variables from $\lbrace \nu, \tau\rbrace \leftrightarrow \lbrace \varphi_8, T\rbrace$, where $\varphi_8$ denotes the 8d dilaton
\begin{equation}\label{eq:8ddilaton}
     e^{-2\varphi_8} = e^{-2\phi} \mathcal{V}_2\, .
\end{equation}
What is important for us is that the \emph{full} U-duality symmetry of the theory is actually larger, namely it consists of $SL(2, \mathbb{Z}) \times SL(3, \mathbb{Z})$, where the modular factor acts solely on the complex structure modulus. In fact, upon introducing the following symmetric  $3\times3$ matrix with unit determinant \cite{Liu:1997mb}
\beq\label{eq:SL3matrix}
 \mathcal{B}= \nu^{1/3} \begin{pmatrix}
		\frac{1}{\tau_2} \quad  \frac{\tau_1}{\tau_2} \quad \frac{c+\tau_1 b}{\tau_2}\\ \frac{\tau_1}{\tau_2} \quad  \frac{|\tau|^2}{\tau_2} \quad \frac{\tau_1 c+|\tau|^2 b}{\tau_2}\\ \frac{c+\tau_1 b}{\tau_2} \quad  \frac{\tau_1 c+|\tau|^2 b}{\tau_2} \quad \frac{1}{\nu} + \frac{|c+\tau b|^2}{\tau_2}
	\end{pmatrix}\, ,
\eeq
which transforms in the adjoint representation of $SL(3, \mathbb{Z})$ (c.f. eq. \eqref{eq:Btransf}), one can rewrite the action \eqref{eq:IIB8d} in a manifestly $SL(2, \mathbb{Z}) \times SL(3, \mathbb{Z})$ invariant way\footnote{Both $SL(2, \mathbb{Z})_{\tau}$ and $SL(2, \mathbb{Z})_{T}$ transformations are embedded within $SL(3, \mathbb{Z})$ as upper and lower block-diagonal subgroups \cite{Kiritsis:1997em}.} 
\begin{equation}\label{eq:IIB8dSL3}
			\begin{aligned}
				S_\text{IIB}^{\text{8d}}\, \supset\, & \frac{1}{2\kappa_{8}^2} \int d^{8}x\sqrt{-g} \left[R -\frac{\partial U \cdot \partial \bar U}{2 U_2^2} + \frac{1}{4} \text{tr} \left( \partial \mathcal{B} \cdot \partial \mathcal{B}^{-1} \right) \right]\, .
			\end{aligned}
\end{equation}
Therefore, we conclude that the moduli space of the theory is described by a coset of the form $\mathcal{M}_{\text{8d}}=SL(2, \mathbb{Z})\backslash SL(2, \mathbb{R})/U(1) \times SL(3, \mathbb{Z})\backslash SL(3, \mathbb{R})/SO(3)$, where the discrete piece corresponds to the U-duality group of the eight-dimensional theory \cite{Hull:1994ys}.
	
In the following, it will be useful to phrase all our discussion using a dual description in terms of M-theory compactified on a three-torus, whose bosonic action reads \cite{Obers:1998fb} 
\begin{align}\label{eq:8d}
	S_\text{M-th}^{\text{8d}}\, \supset\, &\frac{1}{2\kappa_8^2} \int d^{8}x\, \sqrt{-g}\,  \left[ R - \frac{1}{4} \left(g^{i i'} g^{j j'} + \frac{1}{6} g^{i j} g^{i' j'}\right) \partial g_{ij} \partial g_{i' j'} - \frac{1}{2 \mathcal{V}_3^2} \left(\partial C^{(3)}_{123} \right)^2 \right]\, ,
\end{align}
where $g_{ij}$ is the internal metric, $\mathcal{V}_3$ denotes the overall volume in M-theory units and the scalar $C^{(3)}_{123}$ arises by reducing the antisymmetric 3-form field along the $T^3$. In order to make contact with the previous Type IIB perspective, one needs to identify the modulus $U$ with a complex-valued field $\mathcal{T}$, which is defined as follows
\beq\label{eq:T3complexvolume}
\mathcal{T}= C^{(3)}_{123}+ \text{i} \mathcal{V}_3\, ,
\eeq
as well as relate the moduli entering into the matrix $\mathcal{B}$ in eq. \eqref{eq:SL3matrix} with the `unimodular' metric components of the torus, namely $\mathcal{B}_{ij} \rightarrow \tilde{g}_{ij}= \mathcal{V}_3^{-1/3} g_{ij}$. This recovers the action \eqref{eq:IIB8dSL3} written in a manifestly $SL(3, \mathbb{Z})$ invariant form, as explained in detail in Appendix \ref{ap:maxSUGRA8d}.

Once again, one would expect the species scale $\LQG$ to be an automorphic function of the moduli, whose asymptotic behaviour should match the usual species counting procedure, as predicted by \cite{Calderon-Infante:2023ler}. This is summarised in Figure \ref{fig:ch8dgeneric} below, where the convex hull generated by the relevant $\mathcal{Z}$-vectors is explicitly shown.

\begin{figure}[htb]
		\begin{center}
			\subfigure{
				\includegraphics[width=0.45\textwidth]{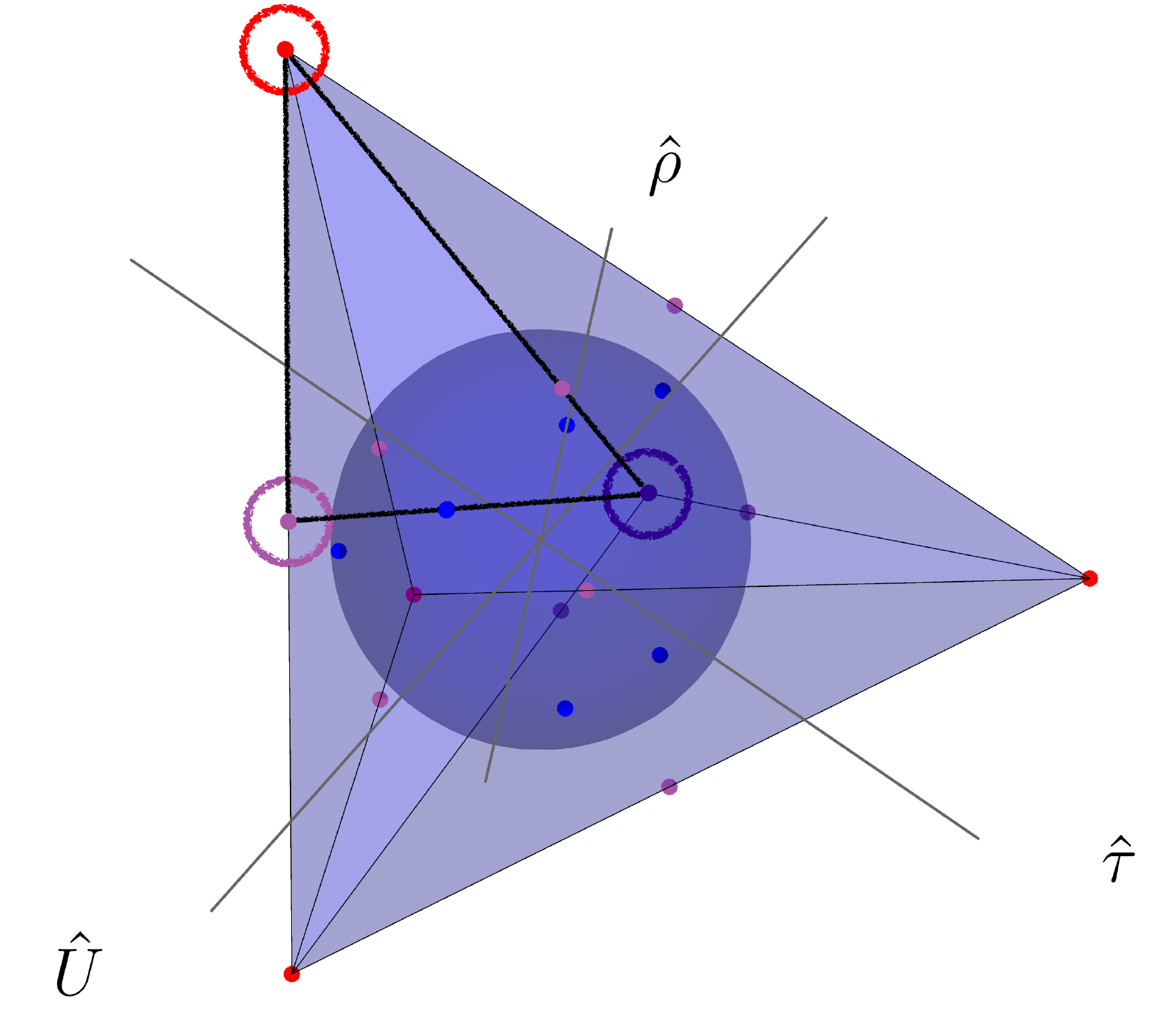}\label{sfig:slice3dCH}
			}
			\subfigure{
				\includegraphics[width=0.45\textwidth]{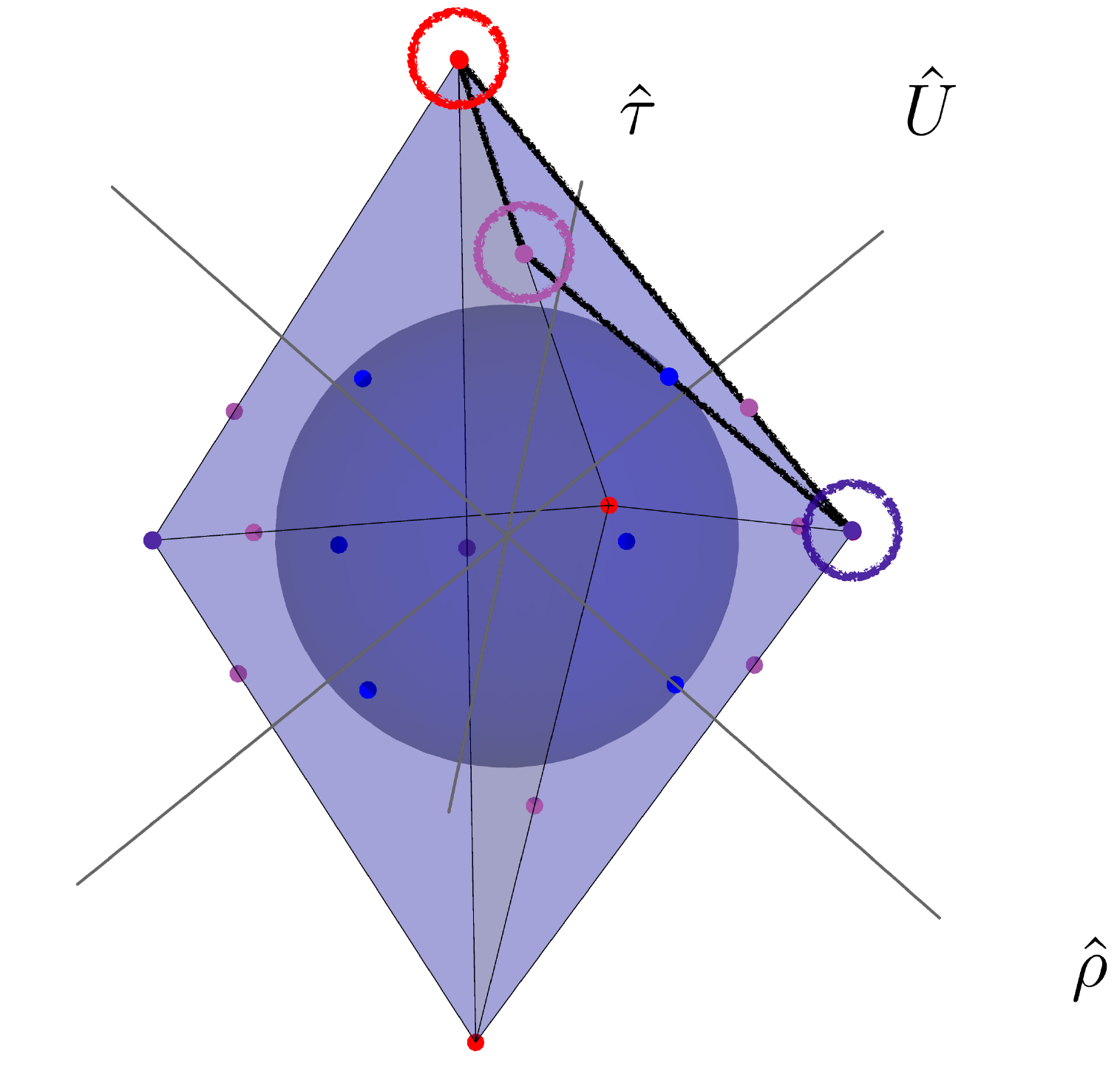}\label{sfig:2ndslice3dCH}
			}
			\caption{\small Convex hull for the species scale in M-theory on $T^3$ ignoring the axionic directions, as seen from two different angles \cite{Calderon-Infante:2023ler}. The blue dots in the facets of the convex hull correspond to single KK towers ($n=1$), the light purple dots in the edges indicate double KK towers ($n=2$) and the purple and red dots in the vertices correspond to triple KK towers ($n=3$) and string towers, respectively. The encircled points represent the particular set of states that we are choosing as representatives of the vertices of the fundamental domain, which is indicated by the triangle tha arises by joining their vertices.}
			\label{fig:ch8dgeneric}
		\end{center}
\end{figure} 


Before proceeding with a systematic study of the higher-dimensional operators appearing in the 8d effective action and their relation to the QG cut-off, let us first try to guess what could be at least the $SL(2, \mathbb{Z})$ dependence of the species scale based on our general discussion from Section \ref{ss:modinvariantss}. There, a natural candidate for a modular invariant species scale function was proposed, which essentially depends on a single real parameter $\gamma$ (c.f. eq. \eqref{eq:lambdaspecies}). In fact, the precise value for $\gamma$ could be easily determined in terms of the spacetime dimension of our theory as well as the nature of the unique infinite distance degeneration in the $SL(2, \mathbb{Z})$ subsector. Therefore, upon applying eqs. \eqref{eq:chargetomassratio} and \eqref{eq:lambdaspecies} in the present eight-dimensional setup for the limit $\mathcal{T} \to \text{i} \infty$, and taking into account that along this limit the 8d theory effectively decompactifies back to the original 11d supergravity, we obtain $\gamma=1$. This leads to the following functional form for (a subset of) the effective number of light species
\beq\label{eq:NspeciesMthy}
N \supset \left[-\text{log} \left(\mathcal{T}_2\,|\eta(\mathcal{T})|^4\right)\right]\, ,
\eeq
which moreover behaves like $N \sim \mathcal{T}_2$ as $\mathcal{T}_2 \to \infty$.

\subsubsection*{The $R^4$ operator}

In order to check the proposed species function \eqref{eq:NspeciesMthy}, as well as to further extend it so as to include the additional $SL(3,\mathbb{R})/SO(3)$ sector, one can look at known higher-dimensional operators appearing in the 8d $\mathcal{N}=2$ effective action. Therefore, we start from the general form for the scalar and gravitational sectors of the theory
\beq
\mathcal{L}_\text{M-th}^{\text{8d}} \supset \frac{1}{2 \kappa_8^2} \left[ \left(R + \sum_k \frac{\mathcal{O}_k (R)}{\LQG^{k-2}}\right) -\frac{\partial \mathcal{T} \cdot \partial \bar{\mathcal{T}}}{2\mathcal{T}_2^2} + \frac{1}{4} \text{tr} \left( \partial \tilde{g} \cdot \partial \tilde{g}^{-1} \right)\right] \ .
\label{eq:scalargrav8dlag}
\eeq
A particularly interesting set of operators within a given spacetime dimension $d$ are those verifying $k=d$, since eqs. \eqref{eq:speciesscale} and \eqref{eq:scalargrav8dlag} predict an asymptotic moduli dependence for them of the form $\LQG^{2-d}\, \mathcal{O}_d (R) \sim N(\vec{\phi})\, \mathcal{O}_d (R)$, where $\vec{\phi}$ denotes collectively all the massless scalars in the theory. In the particular 4d $\mathcal{N}=2$ case analyzed in \cite{vandeHeisteeg:2022btw} (see also Section \ref{s:4dN=2} below) this is precisely the coefficient of the operator $R^2$, which is controlled by the topological genus one partition function $\mathcal{F}_1$. In our 8d setup, it would rather correspond to a term behaving schematically like $R^4$. This was already computed and studied in detail in \cite{Green:1997as,Green:2005ba}, and it reads as follows
\beq
\mathcal{L}^{\text{8d}}_{R^4} = \left( \hat{E}_{3/2}^{sl_3} + 2\hat{E}_{1}^{sl_2} \right)\, t_8 t_8 R^4\ ,
\label{eq:R^4}
\eeq
where $\hat{E}_{1}^{sl_2}, \hat{E}_{3/2}^{sl_3}$ are the regularised Eisenstein series of order 1 and 3/2 for $SL(2,\mathbb{Z})$ and $SL(3,\mathbb{Z})$, respectively (see Appendix \ref{ap:Massform}). In the following, we show that the expression \eqref{eq:R^4} matches asymptotically the diagram depicted in Figure \ref{fig:ch8dgeneric}, including our prediction  \eqref{eq:NspeciesMthy}.

First, notice that due to $SL(2,\mathbb{Z}) \times SL(3,\mathbb{Z})$ duality, it is enough to focus on a fundamental domain containing the minimal non-redundant information captured by the asymptotic convex hull diagram \cite{Calderon-Infante:2023ler, Castellano:2023jjt,Castellano:2023stg}. For concreteness we take the sub-polytope generated by the vertices corresponding to a fundamental Type II string, a 10d Planck scale implementing M/F-theory duality and the 11d Planck mass of the full decompactified theory. Note that written in a particular orthonormal basis (see eq. \eqref{eq:8dchangeofvariables}), this corresponds to the convex cone generated by the following three vectors in Figure \ref{fig:ch8dgeneric}: one of the red dots at the vertices (string), one of the purple dots at the other vertex sharing a common edge (11d Planck mass), and finally one of the light purple dots (10d Planck mass) that belongs to the same facet as the other two and shares an edge with the red dot but not with the purple one. These three vectors comprise the minimal set generating the desired cone representing a fundamental domain. In addition, note that the aforementioned domain automatically includes the vectors associated to yet another partial decompactification of two extra dimensions (light purple dot, 10d Planck mass, at an edge), and one partial decompactification of one extra dimension (blue dot, 9d Planck mass, at the interior of the cone). A particular choice of such representative vectors is given by the encircled dots in Figure \ref{fig:ch8dgeneric}, and the intersection between the fundamental domain and the convex hull is the highlighted triangle that results from joining them. These choice of vectors reads
\begin{equation} \label{eq:Zvectors8d}
	\begin{split} 
		\vec{\mathcal{Z}}_{\text{osc},\, 3} &= \left( 0, -\sqrt{\frac{2}{21}}, \frac{1}{\sqrt{14}} \right) \, , \qquad \vec{\mathcal{Z}}_{\text{KK-M},\, 3} = \left( \frac{1}{4 \sqrt{2}}, -\frac{1}{2 \sqrt{42}}, \frac{1}{4 \sqrt{14}} \right) \, ,\\
		\vec{\mathcal{Z}}_{\text{KK},\, (123)} &= \left( 0, \frac{1}{\sqrt{42}}, \frac{2}{3 \sqrt{14}} \right) \, ,
	\end{split}
\end{equation}
where we follow the notation of ref. \cite{Calderon-Infante:2023ler}, namely $\vec{\mathcal{Z}}=\left(\mathcal{Z}^{\hat \tau}, \mathcal{Z}^{\hat \rho}, \mathcal{Z}^{\hat U}\right)$. In terms of M-theory variables $\lbrace \tau_2, R_3, \mathcal{V}_2 \rbrace$, this fundamental domain is given by the following inequalities
\begin{equation}
    \label{eq:funddomainT3}
\mathcal{V}_2^{9/7} \, R_3^{-2}\,  \geq\, \tau_2\, \geq 1 \, ,  \qquad \mathcal{V}_2 \geq R_3^{-7/6} \, .
\end{equation}
The three corresponding (asymptotic) QG cut-offs associated to the vectors \eqref{eq:Zvectors8d} are given by (in 8d Planck units) 
\begin{equation} \label{eq:QGcutoffs8d}
	\begin{split} 
		m_{\text{str}} &\sim R_3^{1/3}\, \mathcal{V}_2^{-3/14} \, , \qquad M_{\text{pl},\, 10} \sim R_3^{1/12}\, \mathcal{V}_2^{-3/56}\, \tau_2^{-1/8} \, ,\\
		M_{\text{pl},\, 11} &\sim R_3^{-1/6}\, \mathcal{V}_2^{-1/7} \sim \mathcal{V}_3^{-1/6}\, ,
	\end{split}
\end{equation}
which eq. \eqref{eq:R^4} ought to reproduce upon taking appropriate limits within said fundamental domain. 

\underline{\textit{The M-theory regime}}

Let us first consider the region of the selected fundamental domain where $M_{\text{pl},\, 11}$ is the lightest of these three scales, and therefore fixes the species scale, i.e. $\LQG=M_{\text{pl},\, 11}$. This corresponds to the $T^3$ decompactification limit $\mathcal{V}_3 \to \infty$, with $\mathcal{V}_2 \leq R_3^7$, restricted to \eqref{eq:funddomainT3}. This is associated to the $SL(2, \mathbb{Z})$ sector of the theory, since the $\hat{E}_{1}^{sl_2}$ term in \eqref{eq:R^4} dominates in this region. From eq. \eqref{eq:asymptotic behavior} it follows that in all the asymptotic regimes associated to this region we have
\beq
2\hat{E}_{1}^{sl_2}(\mathcal{T}, \bar{\mathcal{T}}) = -2\pi\text{log} \left(\mathcal{T}_2\,|\eta(\mathcal{T})|^4\right)\, \sim\, \frac{\pi^2}{3} \mathcal{V}_3\, \propto\, \left(\frac{M_{\text{pl, 11}}}{M_{\text{pl, 8}}}\right)^{-6}\, ,
\eeq
where the asymptotic dependence should be understood when taking the limit $\mathcal{T}_2\to \infty$. This matches exactly with our prediction in eq. \eqref{eq:NspeciesMthy} for the $SL(2,\mathbb{Z})$ sector, recovering the expected suppression of the $R^4$ term with $\LQG^6$. 

\underline{\textit{The Type II regime}}
 
The regime in which the 10d string scale is the lightest within the fundamental domain \eqref{eq:funddomainT3} is given by $\mathcal{V}_2 \geq R_3^7$, and thus corresponds to $\LQG=m_{\text{str}}$. In this region, the leading contribution to \eqref{eq:R^4} comes from the order-3/2 Eisenstein series, $\hat{E}_{3/2}^{sl_3}$, and all encompassing asymptotic boundaries correspond to the limit $\nu \to \mathcal{V}_2^{-18/7}\, R_3^4\, \tau_2^{-3/2}\to 0$ (where the change of coordinates corresponds to performing Type IIB/M-theory duality in 8d -- see discussion around eq. \eqref{eq:usefulmap}). Thus, upon using the expansion \eqref{eq:Eisenstein3/2}, one finds the leading order contribution
\begin{align}
\label{eq:8dtotypeII}
	\hat{E}_{3/2}^{sl_3}\, \sim\, 2\zeta(3) \mathcal{V}_2^{9/7}\, R_3^{-2}\, \propto\, \left(\frac{m_{\text{str}}}{M_{\text{pl},\, 8}}\right)^{-6}\, ,
\end{align}
which in turn reproduces the expected suppression of the $R^4$ operator with $\LQG^{6}$.

\underline{\textit{Decompactification to 10d}}

Note that the two regimes described so far cover the entire asymptotic region associated to the fundamental domain \eqref{eq:funddomainT3}. However, along certain directions they also coincide with other partial decompactification limits, as presented in the following. Within the selected domain, the region where the 10d Planck mass sets the species corresponds to the boundary where it coincides with the string scale, namely $\tau_2=\, \mathcal{V}_2^{9/7}\, R_3^{-2}$ and $\mathcal{V}_2 \geq R_3^7$. In that case, the mass scale of the double KK tower represented by the light purple dot is lighter than the string scale and we see a decompactification to 10d, with the 10d Planck mass parametrically of the same order as the string scale, namely $\LQG=M_{\text{pl},\, 10}\sim m_{\text{str}}$. The dominant contribution to the $R^4$ term thus takes the same form as in \eqref{eq:8dtotypeII}, which can be equivalently expressed as
\beq
\hat{E}_{3/2}^{sl_3}\, \propto\, \left(\frac{M_{\text{pl},\, 10}}{M_{\text{pl},\, 8}}\right)^{-6} \, ,
\eeq
recovering again the expected result.

\underline{\textit{Decompactification to 9d}}

Finally, let us study the special point corresponding to the center of the facet included in the fundamental domain of Figure \ref{fig:ch8dgeneric}, to which we can associate a single KK tower signaling decompactification from 8d to 9d. This point is such that all the potential candidates for $\LQG$ in \eqref{eq:QGcutoffs8d} scale asymptotically in the same way, namely $\tau_2=\mathcal{V}_2=R_3^7$, which obviously represents a particular geodesic direction towards one of the boundaries of the region defined in \eqref{eq:funddomainT3}. In this limit, we then have $\LQG=M_{\text{pl},\, 9}\sim M_{\text{pl},\, 10} \sim M_{\text{pl},\, 11} \sim m_{\text{str}}$, such that 
\begin{equation}
   \hat{E}_{3/2}^{sl_3} \propto \hat{E}_{1}^{sl_2} \propto  \left(\frac{M_{\text{pl},\, 9}}{M_{\text{pl},\, 8}}\right)^{-6} \, ,
\end{equation}
yielding once again the dependence of the $R^4$ term with the number of species $N=\LQG^{-6}$.



All in all, we conclude that the function $N = \hat{E}_{3/2}^{sl_3} + 2\hat{E}_{1}^{sl_2}$ captures every single relevant asymptotic behaviour of the species scale in 8d maximal supergravity, being moreover invariant under the $SL(2,\mathbb{Z})\times SL(3,\mathbb{Z})$ U-duality group and thus reproducing precisely the convex hull diagram depicted in Figure \ref{fig:ch8dgeneric}.

Finally, as for the previous cases, one can look at other corrections proportional to extra derivatives of the quartic curvature invariants that have already been computed in the literature \cite{Green:2010wi}. Upon doing so, one finds that they are not suppressed by the appropriate power of the species scale along certain type of limits, but by some rather intricate combination of scales. The physical rationale for this discrepancy will be explained later on in Section \ref{ss:gravEFTexpansion}, but for the moment let us stress one more time that we believe the $R^4$-correction to be the one capturing the UV divergence of the eight-dimensional EFT related to the species scale.

\section{Lower dimensional String Theory Examples}
\label{s:4dN=2}

In this section we will focus on 4d $\mathcal{N}=2$ settings arising from Type IIA String Theory compactified on a Calabi--Yau threefold $X_3$. Such theories are known to present, beyond the two-derivative lagrangian, interesting higher-dimensional and higher-curvature corrections. In particular, there is an infinite number of F-terms, which are $\frac{1}{2}$-BPS and thus protected by supersymmetry, ensuring that their dependence with respect to the vector multiplet moduli can be computed exactly. They read as follows \cite{Bershadsky:1993ta, Bershadsky:1993cx,Antoniadis:1993ze,Antoniadis:1995zn}:
\beq
\label{eq:lagrangian}
	\mathcal{L}_\text{IIA}^{\text{4d}} \supset \sum_{g\geq 1}\int d^4\theta\, \mathcal{F}_g (\mathcal{X}^A)\, \mathcal{W}^{2g}\ +\ \text{c.c.}\, ,
\eeq
where $\mathcal{F}_g (\mathcal{X}^A)$ is a chiral superfield that is related to the $g$-loop topological free energy of the supersymmetric closed string, $\theta^{\alpha}$ denote the fermionic $\mathcal{N}=2$ superspace coordinates and $\mathcal{W}_{\mu \nu} = F^+_{\mu \nu} - R^+_{\mu \nu \rho \sigma} \theta \sigma^{\rho \sigma} \theta + \ldots\,$, is the Weyl superfield, which depends (in Euclidean signature) on the self-dual components of the graviphoton field-strength and the Riemann tensor, see e.g. \cite{Antoniadis:1995zn}. Thus, upon performing the integration over the fermionic variables, one obtains terms in \eqref{eq:lagrangian} of the form
\beq
\label{eq:GVterms}
	\mathcal{L}_\text{IIA}^{\text{4d}} \supset \sum_{g\geq 1}\mathcal{F}_g(X^A)\, R_+^2\, F_+^{2g-2}\ +\ \text{c.c.}\, ,
\eeq
where $X^A$ with $A=0,\ldots, h^{1,1}$ denote the bottom (i.e. scalar) components of the chiral superfields $\mathcal{X}^A$.

As originally proposed in \cite{Gopakumar:1998ii,Gopakumar:1998jq}, one can alternatively compute the quantities $\mathcal{F}_g$ for $g\geq 0$ using the duality between Type IIA string theory on $X_3$ and M-theory on $X_3 \times S^1$. For a single BPS particle of mass $m=|Z|$ -- where $Z$ denotes its central charge, one indeed obtains a generating function via a Schwinger-type one-loop computation in the presence of a constant self-dual graviphoton field-strength background, as follows
\begin{align}
\label{eq:generatingseries}
	\notag \sum_{g\geq 0}\mathcal{F}_g\, F_+^{2g-2} &= -\frac{1}{4} \int_0^{\infty}\frac{d\tau}{\tau} \frac{1}{\sin^2{\frac{\tau F_+ \bar Z}{2}}} e^{-\tau m^2}\\
    &= \frac{1}{4} \int_0^{\infty}\frac{d\tau}{\tau} \sum_{g\geq0} \frac{2^{2g} (2g-1)}{(2g)!} (-1)^{g} B_{2g} \left( \frac{\tau F_+}{2}\right)^{2g-2} e^{-\tau Z}\, +\, \mathcal{O}\left(e^{-\frac{Z}{F_+}}\right)\, ,
\end{align}
where in the second step we have changed the integration variable $\tau \to \tau /\bar Z$ and we have performed a perturbative expansion using the following mathematical relation
\begin{align}
	\frac{1}{\sin^{2}(x)} = \sum_{n=0}^{\infty} \frac{2^{2n}(2n-1)}{(2n)!} (-1)^{n-1} B_{2n} x^{2n-2}\, ,
\end{align}
which is valid for $0<|x|<\pi$. Notice that the coupling of the BPS particle to the background field crucially involves the anti-holomorphic piece of the mass \cite{Dedushenko:2014nya}. The $B_{2g}$ are the Bernouilli numbers, which are given by
\begin{align}\label{eq:bernouilli}
	B_{2g}= \frac{(-1)^{g+1} 2 (2g)!}{(2\pi)^{2g}} \zeta(2g)\, .
\end{align}
From eq. \eqref{eq:generatingseries} one may already get a feeling of which $\mathcal{F}_g$ are UV sensitive/divergent versus those which actually provide for a convergent contribution. The claim would be that for $g\geq2$, the above integral converges in the UV, whilst for $g=0,1,$ one needs to adopt some regularization scheme. Indeed, one finds
\begin{align}
\label{eq:divergence}
	\mathcal{F}_g \propto \int_{\varepsilon}^{\infty}d\tau\, \tau^{2g-3}e^{-\tau Z} = Z^{2-2g} \Gamma(2g-2, \varepsilon Z)\, ,
\end{align}
where $\varepsilon$ is nothing but the Schwinger implementation of the UV cut-off \cite{Castellano:2022bvr}. Therefore, for $g > 1$, the incomplete gamma function converges to $\Gamma(2g-2) = (2g-3)!\,$, whilst for the remaining cases one finds a UV divergence that needs to be carefully dealt with.

In what follows, we will study the moduli dependence of the coefficients $\mathcal{F}_g(X^A)$ when probing certain representative infinite distance limits in the vector multiplet moduli space \cite{Lee:2019wij}.\footnote{The quantities $\mathcal{F}_g$, as computed by the topological String Theory, are holomorphic in the chiral coordinates $\mathcal{X}^A$. There exists, however, a holomorphic anomaly in the quantum effective action associated to the contribution of the massless fields \cite{Bershadsky:1993cx}. For our purposes here, it will be enough to focus just on the holomorphic piece.}  We will distinguish between operators that are relevant/marginal (in the Wilsonian sense), from those which are irrelevant (and thus UV convergent). The focus will be placed on understanding whether the EFT expansion proposed in eq. \eqref{eq:gravEFTexpansion} is fulfilled or not in the present setup.

\subsection{The $R^2$ term}
\label{ss:threshold4d}

Let us start with the only relevant BPS operator in the 4d lagrangian \eqref{eq:GVterms}, i.e. the one associated to $g=1$. It is proportional (in Euclidean signature) to the self-dual part of the curvature tensor squared, and its Wilson coefficient can be identified with the A-model topological free energy at genus 1. In addition, the function $\mathcal{F}_1$ has been recently argued to provide for a global definition of the species scale in 4d $\mathcal{N}=2$ theories \cite{vandeHeisteeg:2022btw}.\footnote{The authors in \cite{vandeHeisteeg:2022btw} identify $\mathcal{F}_1$ with the (moduli-dependent) number of species $N$, giving various motivations for this. Note that this is also suggested by the Emergence Proposal \cite{Harlow:2015lma,Grimm:2018ohb,Heidenreich:2017sim,Heidenreich:2018kpg}, since upon taking the species scale as the UV cut-off, the one-loop computation of the $R^2$-term in 4d $\mathcal{N}=2$ theories can be shown to behave asymptotically as $\mathcal{F}_1 \sim \sum_{n=1}^{N} \sim N$ \cite{Castellano:2022bvr}.} 

In order to make contact with our discussion in Section \ref{ss:modinvariantss}, we will henceforth concentrate in theories exhibiting $SL(2,\mathbb{Z})$ invariance. As a simple example, let us consider Type IIA String Theory compactified on the Enriques CY $\left(K3 \times T^2\right)/\mathbb{Z}_2$ \cite{Klemm:2005pd}, which is known to be dual to a heterotic compactification on $K3 \times T^2$. In this case, one finds (at large volume) that the moduli space metric behaves as $G_{T \bar{T}}=\frac{1}{4 T_2^2}\, $, whereas the genus-one topological free energy takes the following simple form \cite{vandeHeisteeg:2023ubh,Grimm:2007tm}
\beq \label{4dtopologicalfreeenergy}
\mathcal{F}_1 =-6 \log \left( T_2 |\eta(T)|^4\right) + \text{const.}\, , 
\eeq
where $T$ denotes the (complexified) K\"ahler modulus associated to the internal torus. In the original Type IIA frame, an emergent string limit arises when taking $T_2 \to \infty$, corresponding to a large volume limit for the internal $T^2$. In the dual heterotic frame, such infinite distance limit is mapped to a perturbative weak coupling point for the fundamental string. Therefore, according to our ansatz \eqref{eq:modspeciesscale}, we predict for the present 4d setup a value $\gamma=1$ (c.f. \eqref{eq:chargetomassratio}), which is precisely what is observed in eq. \eqref{4dtopologicalfreeenergy} above.

An analysis along similar lines can be performed when probing other kind of infinite distance limits within the vector multiplet sector of 4d $\mathcal{N}=2$ theories, including partial decompactifications to M-/F-theory in 5 and 6 spacetime dimensions, respectively. This was done in detail in refs. \cite{vandeHeisteeg:2022btw,vandeHeisteeg:2023ubh}, so we refrain from repeating it here and refer the interested reader to the original works.

\subsection{The UV convergent terms}
\label{ss:UVconvergent}

Let us turn now to the BPS operators in eq. \eqref{eq:GVterms} with $g>1$. Using eq. \eqref{eq:generatingseries}, we find that the contribution to $\mathcal{F}_{g>1}$ due to a particle of mass $m= |Z|$ is
\begin{equation}
\label{eq:Fg>1}
	\mathcal{F}_{g>1}= \frac{(2g-1)}{(2g)!} (-1)^g B_{2g} \int_0^{\infty}d\tau\, \tau^{2g-3}e^{-\tau Z}\, ,
\end{equation}
where one should substitute the appropriate Bernouilli numbers from \eqref{eq:bernouilli}. In the following, we will extract the relevant asymptotic behaviour of these higher-dimensional operators depending on the infinite distance limit that is approached. Later on, in Section \ref{ss:gravEFTexpansion} we will comment on how these examples fit within the general expectation for a gravitational EFT expansion.

\subsubsection*{Large volume limit}

The large volume point (with 4d dilaton fixed and finite) corresponds to a decompactification limit to 5d M-theory, where the M-theory circle grows large. This can be easily understood by looking at the light spectrum of the theory along the aforementioned limit, where it is precisely the tower of D0-branes which become light the fastest (see e.g., \cite{Font:2019cxq}). These states are $\frac12$-BPS, with a mass given by
\begin{equation}
\label{eq:D0mass}
	m_n = 2\pi |n|\, \frac{m_s}{g_s} = |n|\, m_{\rm D0}\, ,
\end{equation}
where $n\in \mathbb{Z} \setminus \{ 0\}$ is the D0-brane charge. Notice that we are excluding the contribution of the massless (i.e. $n=0$) fields here, which would be actually part to the 4d EFT. After substituting in \eqref{eq:Fg>1}, one finds
\begin{align}
\label{eq:Fg>1D0}
	\mathcal{F}^{\rm D0}_{g>1}&= \chi(X_3) \frac{(2g-1) \zeta(2g)}{(2\pi)^{2g}} \sum_{n \in \mathbb{Z} \setminus \{ 0\}}\int_0^{\infty}d\tau\, \tau^{2g-3}e^{-\tau\, n\,  m_{\rm D0}} \notag\\
    &=\chi(X_3) \frac{(2g-1) \zeta(2g)}{(2\pi)^{2g}} \Gamma(2g-2)  m_{\rm D0}^{2-2g}\sum_{n \in \mathbb{Z} \setminus \{ 0\}} \frac{1}{n^{2g-2}} \notag\\
    &= \chi(X_3)\frac{2(2g-1) \zeta(2g) \Gamma(2g-2)}{(2\pi)^{2g}} \frac{\zeta(2g-2)}{m_{\rm D0}^{2g-2}}\, ,
\end{align}
which is of course convergent and moreover depends solely on the mass scale of the tower of states, i.e. $m_{\rm D0}$, instead of the UV cut-off given by the species scale.

\subsubsection*{Decompactification to 6d}

Let us next consider the possibility that our Calabi--Yau threefold presents some elliptic fibration $\pi: X_3 \to B_2$ (for simplicity we assume it to be non-singular). This means, in particular, that there exists an infinite distance limit at large volume within the vector multiplet moduli space, where the base of the fibration blows up, whilst the volume of the elliptic fibre remains constant. Such limit corresponds to a partial decompactification limit to 6d F-theory \cite{Lee:2019wij}, where a tower of bound states with arbitrary D2 and D0-brane charge becomes asymptotically light. The mass spectrum for such tower would read
\begin{equation}
\label{eq:ellipticmass}
	m_{n,\omega} = \frac{2\pi m_s}{g_s} \left| \omega z+n\right|\, ,
\end{equation}
where $z$ is the K\"ahler modulus associated to the elliptic fibre and $(n,\omega) \in \mathbb{Z}^2$ correspond to D0 and D2-brane charge, respectively. To properly account for the effect of such a tower, one needs to sum over the two quantum numbers $(\omega, n)$ independently, yielding
\begin{align}
\label{eq:Fg>1elliptic}
	\mathcal{F}^{\rm ell}_{g>1} (z)&= \chi(X_3)(-1)^{g-1} \frac{(2g-1)B_{2g} \Gamma(2g-2)}{2(2g)!} \frac{1}{m_{\rm D0}^{2g-2}}\sum_{(\omega,n) \in \mathbb{Z} \setminus \{ 0,0 \}} \left(\omega z+n \right)^{2-2g}\notag\\
    &=\chi(X_3)(-1)^{g-1} \frac{(2g-1)B_{2g} \Gamma(2g-2)}{2(2g)!} \frac{1}{m_{\rm D0}^{2g-2}} G_{2g-2}(z)\, ,
\end{align}
where in the last equality we have introduced the holomorphic Eisenstein series 
\begin{align}
\label{eq:holoEisenstein}
	G_{2k}(z) = \sum_{(\omega, n) \in \mathbb{Z}^2 \setminus \lbrace (0,0)\rbrace} \frac{1}{\left(\omega z +n \right)^{2k}}\, .
\end{align}
Notice that the resulting operator is modular invariant,\footnote{The particular case of $g=2$ is a bit subtle, since it appears to be proportional to $G_2(z)$ which by itself is not a modular form (c.f. footnote \ref{fnote:Eisenstein}). However, the holomorphic anomaly \cite{Bershadsky:1993cx} crucially solves this problem by promoting $G_2(z)$ in eq. \eqref{eq:Fg>1elliptic} to its non-holomorphic cousin, namely $\tilde{G}_2 (z, \bar z)= G_2(z)- \frac{\pi}{\text{Im}\, z}$, which now has definite modular weight.} as one can see from the fact that the terms in the lagrangian \eqref{eq:GVterms} are homogeneous in the fields $X^A$ of degree $2-2g$, whose K\"ahler transformation is exactly compensated by that of the graviphoton background $F^+_{\mu \nu}$. Indeed, the relevant $SL(2,\mathbb{Z})$ transformation corresponds to some generalized (double) T-duality which acts on the K\"ahler modulus as follows
\begin{align}\label{eq:Tdualitytrans}
	&z \rightarrow \frac{a\, z + b}{c\, z+d}\,,\qquad \text{with}\ \ \mathcal{A}= \begin{pmatrix}
		a \quad  b\\c \quad  d
	\end{pmatrix} \in SL(2,\mathbb{Z})\, ,
\end{align}
whilst the vectors get transformed linearly through the matrix $\mathcal{A}$, which is moreover embedded into the group of (K\"ahler) symplectic transformations $Sp(2h^{1,1}+2, \mathbb{Z})$, thus leaving the F-terms in \eqref{eq:GVterms} invariant.\footnote{In general, these transformations are more complicated and also take into account the non-trivial fibration structure of the threefold. This involves promoting the (double) T-duality in \eqref{eq:Tdualitytrans} to a Fourier-Mukai transform \cite{Andreas:2004uf, Cota:2019cjx}.} Note that the scale suppressing the tower of BPS operators becomes again that of the D2 particles (equivalently D0 particles, since both have asymptotically the same mass), and not the QG cut-off.

\subsubsection*{Emergent string limits}

Finally, we come to analyze infinite distance points in the vector multiplet moduli space corresponding to emergent string limits. These arise when the Calabi--Yau threefold exhibits some $K3/T^4$-fibration \cite{Lee:2019wij}, with the leading tower of asymptotically light states being the excitation modes of a dual critical string obtained by wrapping an NS5-brane on the generic fibre. As a concrete example, we consider here Type IIA compactified on $\mathbb{P}^{1,1,2,8,12}[24]$, which has $h^{1,1}=3$, $h^{2,1}=243$ and moreover exhibits a $K3$-fibration over a $\mathbb{P}^1$-base. The triple intersection numbers are \cite{Hosono:1994ax}
\beq
	\mathcal{K}_{111}=8\, , \quad \mathcal{K}_{112}=2\, , \quad \mathcal{K}_{113}=4\, , \quad \mathcal{K}_{133}=2,\, \quad \mathcal{K}_{123}=1\, .
\label{eq:triplenumbers}
\eeq
Upon probing the limit $t_b \to \infty$, where we denote by $t_b:= t^2$ the K\"ahler modulus associated to the $\mathbb{P}^1$-base, one encounters an infinite distance boundary of the emergent string kind, as discussed before. There, a critical and asymptotically tensionless string arises by wrapping a NS5-brane on the $K3$-fibre, with tension
\beq
	T_{\rm NS5,\, str} = \Mpf^2 \frac{\mathcal{V}_{K3}}{2\mathcal{V}}\, ,
\label{eq:hettension}
\eeq
where $\mathcal{V}_{K3} = \frac{1}{2} \mathcal{K}_{2ij} t^i t^j = (t^1)^2 + t^1 t^3$ denotes the (classical) volume of the generic fibre and $\mathcal{V}$ that of the threefold. Along the aforementioned limit, the 4d theory admits a dual String Theory interpertation in terms of a perturbative (i.e. weak coupling) limit for an heterotic compactification on $K3 \times T^2$. In such scenario, the quantities $\mathcal{F}_{g>1}$ arise at one-loop level in string perturbation theory \cite{Antoniadis:1993ze,Antoniadis:1995zn}. The dual theory is obtained by compactifying the $E_8 \times E_8$ heterotic string with an $SU(2)$ bundle (of instanton number 12) embedded in each of the $E_8$ factors, and with all non-abelian symmetries higgsed. The remaining $U(1)$ factors therefore come from the 3 vector multiplets associated to the complex dilaton $S= \frac{1}{2}\left(\sigma +ie^{-2\varphi_4}\right)$ \cite{Benmachiche:2008ma}, and the geometric $\{T,U\}$ moduli of the internal torus, as well as the graviphoton. 

Incidentally, the moduli dependence of all the relevant higher-derivative couplings can be encapsulated at once upon defining a generating function $F(\lambda, T, U)= \sum_{g=1}^{\infty} \lambda^{2g} \mathcal{F}^{\rm het}_g(T,U)$. Hence, one finds the following formal expression for $F(\lambda, T, U)$ in the case of interest \cite{Marino:1998pg}
\beq
	F(\lambda, T, U) = \frac{1}{2\pi^2} \int_{\mathscr{F}} \frac{d^2\tau}{\tau_2} \left( \frac{G_4 G_6}{\eta^{24}} \sum_{\Gamma^{2,2}} q^{\frac12 |p_L|^2} \bar{q}^{\frac12 |p_R|^2}\right) \left[ \left( \frac{2\pi i \lambda \eta^3}{\theta_1(\tilde \lambda| \tau)}\right)^2 e^{-\frac{\pi \tilde \lambda^2}{\tau_2}}\right]\, .
\label{eq:Fheteroticoneloop}
\eeq
Here $\mathscr{F}$ denotes the $SL(2,\mathbb{Z})$ fundamental domain, $G_4(\tau)$ and $G_6(\tau)$ are holomorphic Eisenstein series (c.f. eq. \eqref{eq:holoEisenstein}), $\theta_1$ is the Jacobi theta function with characteristics $(1/2,1/2)$, and we have defined the quantities
\beq
	\tilde \lambda = \frac{p_R \tau_2 \lambda}{\sqrt{2T_2 U_2}}\, , \qquad q=e^{2\pi i\tau}\, ,
\eeq
where $p_{L,R}$ are the right-/left-moving momenta along the torus: 
\begin{align}
   p_L&= \frac{1}{\sqrt{2T_2U_2}} \left( n_1 +n_2 \bar{T} + m_2 U +m_1 \bar{T}U\right)\, ,\notag \\
   p_R&= \frac{1}{\sqrt{2T_2U_2}} \left( n_1 +n_2 T + m_2 U +m_1 T U\right)\, .
\end{align}
The important point for us is that all Wilsonian couplings captured by $F(\lambda, T, U)$ arise at one-loop and are thus proportional to $(S-\bar S)^0$ in the string frame. Therefore, upon going to the 4d Einstein frame and taking the perturbative limit, namely when $S \to i \infty$ -- and for generic values of the moduli $\{ T,U\}$, the functions $\mathcal{F}_{g>1}$ behave as follows
\beq
	\mathcal{F}^{\rm het}_{g>1} \sim \Mpf^{2-2g}(\text{Im}\, S)^{g-1} \sim \left(m_{\text{str, het}}\right)^{2-2g}\, ,
\eeq
where $\Mpf$ is the 4d Planck scale and $m_{\text{str, het}}$ denotes the fundamental heterotic string scale. In the original type IIA frame, this means that upon probing the limit $t_b \to \infty$, one finds
\beq \label{eq:F_g>1emergentstring}
	\mathcal{F}^{\rm IIA}_{g>1} \sim \Mpf^{2-2g} t_b^{g-1} \sim (T_{\rm NS5, \, str})^{1-g}\, ,
\eeq
with $T_{\rm NS5,\, str}$ given in eq. \eqref{eq:hettension} above. In this case, the scale suppressing the F-terms \eqref{eq:GVterms} does coincide with the species scale along the limit $t_b \to \infty$, since the latter is given by the (emergent) heterotic string scale. However, as we comment further in Section \ref{ss:gravEFTexpansion} below, the power does not agree with the expectations from eq. \eqref{eq:gravEFTexpansion}.

\subsection{General Lessons}
\label{ss:gravEFTexpansion}

After the previous general considerations, we are now in a position to discuss the asymptotic mass dependence of the irrelevant F-terms appearing in generic 4d $\mathcal{N}=2$ theories arising from quantum gravity. We argue in the following that the relevant UV scale controlling these terms in the EFT is the mass scale of the lightest tower (rather than the species scale), such that the series expansion reads
\beq
\mathcal{L}_{\mathrm{EFT,\,}4} \supset \sum_{n > 4} \frac{\mathcal{O}_n (R)}{m_{\rm tower}^{n-4}} \, ,
\label{eq:gravEFTexpansionIII}
\eeq
where $n$ denotes the mass dimension of the corresponding operator. To see this, we first note that the dimension of the operators $R_+^2\, F_+^{2g-2}$ is $n=2g+2 > 4$ for $g>1$, such that they are all irrelevant in the Wilsonian sense. Furthermore, the asymptotic moduli dependence arising in the three possible types of infinite distance boundaries within the vector multiplet moduli space matches the behaviour exhibited in \eqref{eq:gravEFTexpansionIII}, where $m_{\rm tower}$ becomes either $m_{\rm D0}$ or $T_{\rm NS5,\, str}^{1/2}$. For instance, in the large volume limit we found $\mathcal{F}_{g>1} \propto m_{\rm D0}^{2-2g}$, c.f. eq. \eqref{eq:Fg>1D0}, whilst for the emergent heterotic string the dependence was of the form $\mathcal{F}_{g>1} \propto \left(T_{\rm NS5,\, str}^{1/2} \right)^{2-2g}$, see eq. \eqref{eq:F_g>1emergentstring}.

Our aim in this section will be to take the first steps towards understanding whether this observed suppression of the gravitational irrelevant terms by the lightest scale of the tower is an artifact due to some sort of fine tuning \cite{Heckman:2019bzm}, or rather it represents some general behaviour in quantum gravity. We distinguish between decompactification and emergent string limits in what follows, for reasons that will become clear along the way.

\subsubsection*{Decompactification Limits}

Let us consider first infinite distance limits signalling towards decompactification of one or more internal dimensions. Along these, the dominant tower of states becomes the Kaluza-Klein replica, whose masses scale inversely with the volume of the decompactifying cycle.

In the supersymmetric examples studied both in Sections \ref{s:Examples} and \ref{ss:UVconvergent} it was found that for such decompactification limits, the suppression exhibited by those operators which are \emph{irrelevant} (both in the lower and higher-dimensional theory) was of the form
\beq
\mathcal{L}_{\mathrm{EFT,\,}d} \supset \sum_{n > d} \frac{\mathcal{O}_n (R)}{\Lambda_{\rm UV}^{n-d}} \, ,
\label{eq:gravEFTexpansionII}
\eeq
which follows the usual rules of EFT expansions with $\Lambda_{\rm UV}=m_{\rm KK} \ll \LQG$. This includes the BPS operators $\partial^4 R^4$ and $\partial^6 R^4$ appearing in maximally supersymmetric setups in $8\leq d \leq 10$, as well as those of the form $R_+^2\, F_+^{2g-2}$ (for $g>1$) arising in 4d $\mathcal{N}=2$ theories. What we would like to point out here is that this behaviour would fit into a general picture where \emph{gravitational} operators which are relevant/marginal are suppressed by the species cut-off (with respect to the Einstein-Hilbert term), as in eq. \eqref{eq:gravEFTexpansion}, whilst those which are irrelevant follow the pattern in \eqref{eq:gravEFTexpansionII}. A potential heuristic explanation would be the following: The former probe the UV physics associated to quantum gravity and therefore should be controlled by the corresponding QG cut-off. The latter, on the other hand, are typically convergent in the UV, and thus end up being controlled by the mass scale of the towers/states that have been integrated out.

To motivate this claim, let us consider the following simple example: $k$-(super)graviton scattering in $d$ spacetime dimensions. We focus here on the contributions arising from $\frac12$-BPS operators of the schematic form $R^k$. In fact, the contribution of a KK-like tower to such an amplitude can be nicely expressed using the worldline formalism. Therefore, assuming that the vertex insertions saturate alone the fermionic integrals over the zero modes \cite{Green:1999by}, one finds
\beq
\mathcal{A}_{k,\,d}\, =\, \tilde K \sum_{n \in \mathbb{Z} \setminus \{ 0\}} \int d^dp \int_{0}^{\infty} \frac{d\tau}{\tau}\, \tau^k\, e^{-\tau \left( p^2+m_n^2\right)}\, ,
\label{eq:kgravscattering}
\eeq
where $m_n= |n|\, m_{\rm KK}$ and $\tilde K$ denotes the appropriate kinematic factor that accounts for the Lorentz structure of the amplitude.\footnote{The kinematic factor $\tilde K$ can be easily obtained as a linearized version of the relevant operator whose corrections we are computing (e.g., $R^k$).} Upon performing the integration over the loop momentum variables one is left with
\beq\label{eq:kgravscatteringKKtower}
\mathcal{A}_{k,\,d}\, =\, \tilde K \sum_{n \in \mathbb{Z} \setminus \{ 0\}} \int_{0}^{\infty} d\tau\, \tau^{k-1-d/2}\, e^{-\tau\, m_n^2}\, .
\eeq
Notice then, that the integral converges in the UV (i.e. $\tau \sim 0$ region) if and only if $2k-d >0$, which is equivalent to ask for the operator $R^k$ to be irrelevant --- in the Wilsonian sense --- in the $d$-dimensional theory. Still, this does not guarantee the full amplitude to be convergent. To ensure this, one needs to subsequently perform the summation over all KK quantum numbers, thus yielding
\beq\label{eq:kgravscatteringKKtowerII}
\mathcal{A}_{k,\,d}\, =\, 2\tilde K\, \Gamma\left(k-\frac{d}{2}\right) \sum_{n >0} \frac{1}{m_n^{2k-d}} = 2\tilde K\, \Gamma\left(k-\frac{d}{2}\right) \frac{\zeta(2k-d)}{m_{\rm KK}^{2k-d}}\, .
\eeq
The resulting amplitude is thus finite if the stronger condition, $2k-(d+1)>0$, is fullfilled, due to the convergence properties of Riemann's $\zeta$-function. Notice that this is equivalent to the operator being irrelevant in the $D=(d+1)$ parent theory, in agreement with the discussion around eq. \eqref{eq:gravEFTexpansionII}. As a side remark, notice that if instead of performing the summation over all KK modes we instead truncate up to some maximum excitation number $N_{\rm max}$ --- such that $N_{\rm max} m_{\rm KK} \sim \LQG$, one still recovers the above result up to a correction of the form $\mathcal{A}^{\rm corr}_{k,\,d} \sim \Mpd^{d-2}\,\LQG^{2-2k}$, which despite being similar to the expectation in \eqref{eq:gravEFTexpansion}, is of course subleading with respect to \eqref{eq:kgravscatteringKKtowerII}.

In a similar vein, one could consider operators involving spacetime derivatives of the Riemann tensor, such as $\partial^{2\ell} R^k$, whose mass dimension is $2(\ell+k)$. Indeed, essentially the same argument as before leads to the following schematic form of the amplitude (after integrating over loop momenta)
\beq
\mathcal{A}_{k + \ell,\,d}\, =\, \tilde K' \sum_{n \in \mathbb{Z} \setminus \{ 0\}} \int_{0}^{\infty} d\tau\, \tau^{k+ \ell-1-d/2}\, e^{-\tau\, m_n^2} = 2\tilde K'\, \Gamma\left(k+\ell-\frac{d}{2}\right) \frac{\zeta(2(k+\ell)-d)}{m_{\rm KK}^{2(k+\ell)-d}}\, ,
\label{eq:kgravscatteringwithlderivatives}
\eeq
whose convergence properties now depend on whether $2(k+\ell)-(d+1) >0$ or not, which again is equivalent to the corresponding operator being irrelevant or otherwise.

For completeness, let us mention that the above analysis may be easily generalized to the case of $p$ compact extra dimensions. Indeed, restricting ourselves for simplicity to the case of a $p$-dimensional flat torus with metric $G_{ij}$, the one-loop contribution to the $k$-point supergraviton amplitude in $d=D-p$ dimensions can be computed to be
\begin{align}
\mathcal{A}_{k,\,d}\, &=\, \tilde K\, \sum_{\lbrace n_i\rbrace}\, \int_{0}^{\infty} d\tau\, \tau^{k-1-d/2}\, e^{-\tau\,G^{ij} n_i n_j}\notag\\
&=\, \tilde K\, \left(\det G_{ij}\right)^{1/2}\,\sum_{\lbrace \ell_i\rbrace}\, \int_{0}^{\infty} d \tilde{\tau}\, \tilde{\tau}^{-k-1+d/2+p/2}\, e^{-\tilde{\tau}\,G_{ij} \ell^i \ell^j}\, ,
\end{align}
where in the second step we have performed a Poisson resummation, thus exchanging the sum over momenta $\lbrace n_i\rbrace$ with that of (worldline) windings $\lbrace \ell_i\rbrace$ and we have defined a new variable $\tilde{\tau}=\tau^{-1}$. Notice that the UV contribution (i.e. $\tilde{\tau} \to \infty$) is now associated to the zero winding sector, and its convergence depends on whether the corresponding higher-curvature operator --- of dimension $2k$ --- is irrelevant in the parent $D$-dimensional theory, namely whether $d+p \leq 2k$ is fulfilled or not.

\subsubsection*{Emergent string limits}

Next, we turn to emergent string limits, where the dominant tower of asymptotically light states becomes the excitation modes of a (possibly dual) fundamental string. Recall that it was precisely along these limits where the EFT expansion in \eqref{eq:gravEFTexpansion} was fulfilled in maximal supergravity setups, even for irrelevant operators such as $\partial^4 R^4$ and $\partial^6 R^4$ (c.f. Section \ref{s:Examples}).

A first clue that the situation is dramatically different for string towers than for KK towers is provided by the (asymptotic) degeneracy of states. Namely, even if the one-loop contribution to say $\mathcal{A}_{k,\,d}$ converges in the UV for each oscillator mode in the tower (whenever the corresponding operator is irrelevant, as discussed after \eqref{eq:kgravscatteringKKtower}), the summation over the oscillator number $n$ diverges, a priori
\beq
\mathcal{A}^{\rm osc}_{k,\,d} \propto \tilde K\, \Gamma\left(k-\frac{d}{2}\right) \sum_{n >1} \frac{d_n}{m_n^{2k-d}} \to \infty\, ,
\eeq
where $m_n^2= 2\pi m_s^2 (n-1)$ and $d_n \sim n^{-11/2} e^{4\pi \sqrt{2n}}$ in the case of a e.g., Type II fundamental string. This means, in particular, that a simple analysis using worldline/QFT techniques is not valid anymore when probing the (strict) weak coupling point of a critical string, as we already knew. Instead, a worldsheet computation should be more appropriate. In this regard, what does string perturbation theory tell us about the suppression of these terms in a generic gravitational EFT action?  

To address this question, let us come back at the 4d example from Section \ref{ss:UVconvergent}, where we considered Type IIA String Theory on $\mathbb{P}^{1,1,2,8,12}[24]$. From this simple example we can extract two important lessons.  First, even for emergent string limits, one can in principle find an \emph{infinite} number of gravitational terms in the effective action which do not accommodate the ansatz proposed in \eqref{eq:scalargravDlag}. This follows from a straightforward genus counting, such that whenever the higher-curvature operator in question receives a \emph{tree-level} contribution in string perturbation theory, the behaviour predicted by \eqref{eq:scalargravDlag} will be automatically fulfilled. On the contrary, if the leading term --- whenever $g_s^{-1} \gg 1$ --- comes at \emph{one-loop} order, one rather finds agreement with eq. \eqref{eq:gravEFTexpansionIII}, thus giving some a priori smaller suppression. 
However, this does not cause any real problem though, since the scale controlling these operators relative to the Einstein-Hilbert term behaves like $m_s^{\frac{n-d}{n-2}}$, such that for large enough classical dimension $n$ it asymptotes to $m_s$, which precisely provides for the species cut-off $\Lambda_{\rm QG}$.


On hindsight, the reason why the higher-dimensional examples analyzed in Section \ref{s:Examples} presented the `correct' dependence with the string mass was that all such operators exhibited some non-vanishing contribution at tree-level. Thus, according to the discussion in the previous paragraph, one would have indeed expected precise agreement with the expansion in eq. \eqref{eq:gravEFTexpansion}. Analogously, one can check that in the 4d $\mathcal{N}=2$ example above, when viewed from the heterotic frame, the only higher-derivative F-term of the form \eqref{eq:lagrangian} that receives a tree-level contribution is the $R^2$-operator, which is the relevant one (again in the Wilsonian sense) and moreover can be seen to descend from the 10d $R^4$-interaction via classical reduction. 

\section{Summary and Discussion}
\label{s:discussion}

In the present work we have pursued the idea of trying to identify the species scale as the energy cut-off signalling the breakdown of any semi-classical description of gravity \cite{Dvali:2007hz,Dvali:2007wp}. The main focus has been placed on looking at the moduli dependence of certain higher-dimensional operators in the EFT expansion, following the ideas put forward in \cite{vandeHeisteeg:2022btw,vandeHeisteeg:2023ubh}. Such terms seem to probe the ultra-violet nature of quantum gravity in a non-trivial way, and indeed can have a great impact on the IR physics through e.g., (small) Black Holes and entropy considerations (see e.g., \cite{Calderon-Infante:2023uhz,Cribiori:2023ffn}). 

We thus first described what are the expectations from a general EFT analysis in the presence of gravity in Section \ref{ss:basics}.
Next, in Section \ref{ss:modinvariantss} and based on duality symmetry considerations, we proposed a strategy to look for potential candidates for a globally-defined species scale function in certain instances (see also \cite{Cribiori:2023sch} for related ideas). We exemplified such considerations within the particular case of $SL(2,\mathbb{Z})$, which arises in many setups in String Theory and where the mathematical theory of modular forms is fully developed, such that it can be readily applied. However, we noticed that the minimal physical requirements for a bona-fide species scale, namely the fact that it should be bounded from above --- by the Planck scale --- and that it should go to zero asymptotically, are not strong enough so as to uniquely determine such functions. These can be given in general by combinations of `well-behaved' automorphic forms (see Appendix \ref{ap:Massform}), such that there is no fundamental reason a priori to select any specific one. Despite this apparent freedom, and based on our String Theory experience as well as certain threshold computations existing in the literature \cite{Kaplunovsky_1995}, we proposed an educated guess in terms of some particular (regularised) non-holomorphic Eisenstein function as a proxy for the species scale in any $SL(2,\mathbb{Z})$-sector of the theory. Note that, even though the rest of the analysis of higher curvature corrections carried out in this paper is independent of the argument for the proposed general form of the species scale, we also showed both qualitative and quantitative agreement in all the cases where the duality group includes an $SL(2,\mathbb{Z})$-sector.

One of the main questions motivating this work was to explore to what extent the formulation of the species scale as a quantity defined over the whole moduli space from analyzing (protected) higher curvature corrections, originally proposed in the context of 4d $\mathcal{N}=2$ theories \cite{vandeHeisteeg:2022btw}, could be promoted to a general prescription that recovers the right asymptotic results.  In order to address this, we tested these ideas in well-understood, highly supersymmetric setups. In particular, we analysed the structure of the first higher-curvature BPS-protected operators in maximal supergravity in 10, 9 and 8 dimensions, as well as revisited the 4d $\mathcal{N}=2$ scenario arising from e.g., Type II compactifications on Calabi--Yau manifolds \cite{vandeHeisteeg:2022btw}. For the specific case of maximal supergravity in $d=8,9,10$, the first non-trivial correction to the two-derivative lagrangian, arises from certain combinations of four Riemann tensors (see Section \ref{s:Examples}). In all these cases, the moduli-dependent function controlling such terms in the EFT was seen to match asymptotically with the usual species counting computations \cite{Castellano:2022bvr, Calderon-Infante:2023ler} in a non-trivial fashion. This means that, starting from any dual frame, one can retrieve the fundamental QG scale at any other corner of the duality web of the theory (including those living in higher dimensions) by looking at the moduli dependence of the corresponding Wilson `coefficient'. This extends the results from refs. \cite{vandeHeisteeg:2022btw,vandeHeisteeg:2023ubh}, which we reviewed and broadened in Section \ref{s:4dN=2}. There, a similar procedure was performed in 4d theories preserving eight supercharges. Despite this encouraging agreement with our expectations from Section \ref{ss:basics}, several relevant observations regarding the scale(s) suppressing the set of higher-curvature corrections were made. First, the simple expansion in power series of the inverse cut-off was seen to arise only \emph{asymptotically}. Therefore, even though when expanding around any infinite distance corner of the moduli space the different gravitational operators are suppressed by the appropriate species scale, this may be no longer the case when venturing towards the bulk. We believe this has to do with the fact that the expansion in \eqref{eq:gravEFTexpansion} should be taken as some sort of approximation strictly valid close to the infinite distance boundaries, where weak couplings appear and the perturbative series rapidly converge. There, the classical dimension of the different operators in the EFT expansion is a good proxy, whilst in the interior, large quantum corrections (both perturbative and non-perturbative) may occur, thus giving rise to big `anomalous dimensions'. Second, we found that certain non-leading \emph{gravitational} operators in the theory, given by higher derivatives of four Riemann tensors in maximal supergravity \cite{Green:2010kv,Green:2010wi} or higher-dimensional F-terms of the form $R^2\, F^{2g-2}$ in the 4d setup ($F_{\mu \nu}$ being the graviphoton field strength), do not seem to exactly follow the pattern proposed in eq. \eqref{eq:gravEFTexpansion}, but are instead controlled by different integer powers of the lightest tower scale (rather than the species scale). This was interpreted as pointing out that such operators are not really sensitive to the UV cut-off of the theory, but instead are generated by integrating out the towers in a UV-convergent way. Therefore, it seems reasonable that the right way to proceed in order to properly identify the species scale in a given theory is to look at the low-dimensional (but absent in the two-derivative action) relevant or marginal gravitational operators, which really probe the UV-physics of the theory. 

Another interesting feature that we observed was that these operators present a particular moduli dependence tightly related to the moduli space geometry. Indeed, the functional coefficients were seen to satisfy certain eigenvalue/Poisson equations with respect to the moduli space laplacian. In some cases, this can be related to supersymmetry (see e.g., \cite{Green:1998by,DHoker:2022dxx}), but it is not entirely clear if it is really a consequence of the latter, of rather some deep property of the species scale. In particular, it would be interesting to investigate whether the presence of this eigenvalue/Poisson equation could be related to certain IR Swampland conditions, such as the Scalar Weak Gravity Conjecture \cite{Palti:2017elp} (see also \cite{Lee:2018spm,Gonzalo:2020kke}). We leave further investigation of these issues for future work.

Finally, several questions can be raised at this point. First of all, the role of the species scale in the gravitational EFT expansion might not be fully understood yet. It would be interesting to pinpoint precisely which operators can be expected to be suppressed by the appropriate powers of the species scale (instead of just that of the towers). One could also try to extend this analysis to other setups both in different number of spacetime dimensions and supersymmetries. This is particularly interesting in the case of 16 supercharges or less, where certain infinite distance limits probe running-solutions (i.e., not vacua) \cite{Etheredge:2023odp} and where the computation of the species scale seems challenging \cite{Castellano:2023jjt,Castellano:2023stg}. Finally, one could try to study further implications of these considerations within the Swampland program, as well as to revisit certain naturalness/fine tuning arguments that are commonly employed in model building scenarios.
\newpage

\vspace{1.5cm}
\centerline{\bf Acknowledgments}
\vspace{0cm}
		
We would like to thank F. Marchesano, M. Montero, D. Prieto, I. Ruiz, A. Uranga,  I. Valenzuela  for useful discussions and correspondence. A.C. would like to thank the Theoretical Physics Department at CERN for hospitality and support, where most of the work was completed. This work is supported through the grants CEX2020-001007-S and PID2021-123017NB-I00, funded by MCIN/AEI/10.13039/501100011033 and by ERDF A way of making Europe.  The work of A.C. is supported by the Spanish FPI grant No. PRE2019-089790 and by the Spanish Science and Innovation Ministry through a grant for postgraduate students in the Residencia de Estudiantes del CSIC.


\begin{appendix}

\section{Relevant Automorphic Forms}
\label{ap:Massform}

This appendix serves as a mathematical compendium for the relevant set of automorphic functions that appear at several instances in the discussion from the main text. We will focus mostly on the discrete groups $SL(2, \mathbb{Z})$ and $SL(3, \mathbb{Z})$, since they capture the duality symmetries in 10d, 9d and 8d maximal supergravity (see Section \ref{s:Examples}). A similar analysis can be performed for the (bigger) duality groups that arise upon reducing the number of non-compact spacetime dimensions, but we refrain from reviewing those in the present work.

\subsection{$SL(2, \mathbb{Z})$ Maas waveforms}

An automorphic function of a given Lie group $G$ is defined as a map from a space $\mathcal{M}$ admitting some natural group action for $G$, to $\mathbb{R}$ (or more generally $\mathbb{C}$), such that it is left invariant under the corresponding group action for any $g \in G$. In the following, we will particularise to those automorphic functions of $SL(2, \mathbb{Z})$ which are moreover real analytic, since they may appear as (generalised) `Wilson coefficients' in the EFT expansion of our gravitational effective field theories. In fact, there is an economic way to generate such analytic functions as eigenfunctions of some appropriate elliptic operator previously defined. Now, since we want these functions to be automorphic forms as well, we can simply take the hyperbolic Laplace operator, which is both elliptic and $SL(2,\mathbb{Z})$-invariant (in a precise sense that we specify below). This operator reads
\beq\label{eq:SL2Laplacian}
\Delta_2 = \tau_2^2 \left( \frac{\partial^2}{\partial \tau_1^2} + \frac{\partial^2}{\partial \tau_2^2}\right)\, ,
\eeq
where again we define $\tau=\tau_1 + \text{i} \tau_2$. Note that this is nothing but the Laplacian operator associated to the metric \eqref{eq:scalarlag}. Therefore, the eigenfunctions of this operator --- which are moreover modular invariant --- are called singular Maas forms \cite{DHoker:2022dxx}. Here we will be interested, for reasons that will become clear later on, in a subgroup of such set of functions, those denoted simply as Maas forms, which have the additional property of growing polynomially (instead of exponentially) with $\tau_2$, as $\tau_2 \to \infty$. An example of Maas form that plays a key role in the discussion from the main text are the so-called non-holomorphic Eisenstein series \cite{DHoker:2022dxx}
\beq\label{eq:nonholoEisenstein}
\frac{\Gamma(\ell)}{2\pi^{\ell}}\, E_{\ell}^{sl_2}(\tau, \bar \tau) \equiv \pi^{-\ell}\, \Gamma(\ell)\, \frac{1}{2} \sum_{(m, n) \in \mathbb{Z}^2 \setminus \lbrace (0,0) \rbrace} \frac{\tau_2^\ell}{\left| m+n\tau\right|^{2\ell}}\, ,
\eeq
which converge absolutely if $\text{Re}\, \ell >1$. It can be shown (upon using that the fractional linear transformation in eq. \eqref{eq:modtransf} conmutes with the operator $\Delta_2$), that indeed $E_{\ell}^{sl_2}(\tau)$ are both automorphic and eigenfunctions of the hyperbolic Laplacian, with eigenvalue given by $\ell(\ell-1)$. The polynomial growth of the Eisenstein series can be also easily understood, since upon taking the limit $\tau_2 \to \infty$, the infinite series is clearly dominated by the terms with $n=0$, which grows as $\tau_2^\ell$. More precisely, the functions $E_{\ell}^{sl_2}(\tau)$ have an alternative Fourier expansion in $\tau_1$, which can be obtained upon Poisson resumming\footnote{\label{fnote:Poissonresummation}The Poisson resummation identity reads as follows \begin{equation}
    \notag \sum_{n \in \mathbb{Z}} F(x+na) = \frac{1}{a} \sum_{k \in \mathbb{Z}} \tilde{F} \left(\frac{2\pi k}{a} \right) e^{2\pi i k x/a}\, ,
\end{equation}
with $\tilde{F} (\omega)=\int_{-\infty}^{\infty} dx\, F(x) e^{-i \omega x}$ the Fourier transform of $F(x)$.} on the integer $n$, yielding
\begin{align}\label{eq:nonpertexpansion}
	\notag E_{\ell}^{sl_2} =\, & \bigg[ 2\zeta(2\ell) \tau_2^{\ell} + 2\pi^{1/2}\frac{\Gamma(\ell-1/2)}{\Gamma(\ell)} \zeta(2\ell-1) \tau_2^{1-\ell}\\
 &+ \frac{8 \pi^\ell \tau_2^{1/2}}{\Gamma(\ell)} \sum_{m=1}^{\infty} m^{\ell-1/2} \sigma_{1-2\ell} (m)\, \cos(2\pi m \tau_1)\, K_{\ell-1/2} (2\pi m \tau_2)\bigg]\, ,
\end{align}
where $\sigma_{1-2\ell} (m) = \sum_{d|m} d^\ell$ runs over all divisors $d$ of $m$, and $K_\ell(y)$ is the modified Bessel function of second kind, which is defined as follows
\begin{equation}
    K_\ell(y)=\frac{1}{2} \int_0^{\infty} dx\, x^{\ell-1} \exp \left[ -\frac{y}{2} \left( x + \frac{1}{x}\right)\right]\, ,
    \end{equation}
and decays asymptotically as $ K_\ell(y) \sim \sqrt{\frac{\pi}{2y}} e^{-y}$ for $y \to \infty$.

Let us finally mention that the modular function $\left(2\pi^{\ell}\right)^{-1} \Gamma(\ell) E_{\ell}^{sl_2}(\tau)$, when seen as a function also of the variable $\ell$, has a meromorphic continuation to al $\ell\in \mathbb{C}$, which is thus analytic everywhere except for simple poles at $\ell=0,1$. Moreover, if the divergence for $\ell=1$ is `extracted', namely upon selecting the constant term (with respect to $\ell$) in the Laurent series for $E_{\ell}^{sl_2}$ at $\ell=1$, one obtains the following function \cite{DHoker:2022dxx}
\beq
2\pi\left(\gamma_{\text{e}} - \log 2\right) - \pi \log \left( \tau_2\,|\eta(\tau)|^4 \right)\, ,
\eeq
where $\gamma_{\text{e}}$ is the Euler-Mascheroni constant and $\eta(\tau)$ denotes the Dedekind eta function, which may be defined as
\beq \label{eq:Dedekind}
\eta(\tau) = q^{\frac{1}{24}} \prod_{k=1}^{\infty} \left( 1-q^k\right)\, , \qquad q=e^{2\pi i \tau}\, .
\eeq
To conclude, let us note that even though the function $\hat{E}_{1}^{sl_2}(\tau) = - \pi\log \left( \tau_2\,|\eta(\tau)|^4 \right)$ arises as some sort of analytic extension of $E_{1}^{sl_2}(\tau)$, it is actually not a Maas form, since $\Delta_2 \hat{E}_{1}^{sl_2}(\tau)$ is not proportional to $\hat{E}_{1}^{sl_2}(\tau)$ itself but it rather gives a constant value. This can be easily checked upon noting that $\partial \bar \partial \hat{E}_{1}^{sl_2}(\tau)= \frac{\pi}{4 \tau^2_2}$, as well as $\Delta_2=4 \tau_2^2 \partial \bar \partial$, where we have defined $\partial = \partial/\partial \tau$ and $\bar \partial = \partial/\partial \bar \tau$. In any event, what remains true is that the large modulus behaviour of $\hat{E}_{1}^{sl_2}(\tau)$ matches with that expected for $E_{\ell=1}^{sl_2}(\tau)$, since upon using the Fourier series expansion for $\eta(\tau)$
\beq
\eta(\tau) = q^{\frac{1}{24}} \left( 1-q-q^2+q^5 + \mathcal{O}(q^7) \right)\, ,
\eeq
one finds the following relevant asymptotic expression 
\beq \label{eq:asymptotic behavior}
-\pi \text{log} \left(\tau_2\,|\eta(\tau)|^4\right)\, \sim\, -\pi \text{log} \left(\tau_2\,e^{-\frac{\pi \tau_2}{3}}\right)\, \sim\, \frac{\pi^2}{3} \tau_2 - \pi \text{log} (\tau_2)\, ,
\eeq
whose first term precisely is $2 \zeta(2) \tau_2$ (c.f. eq. \eqref{eq:nonpertexpansion}).

\subsection{$SL(3, \mathbb{Z})$ Maas waveforms}

Motivated by the examples considered in the main text --- in particular maximal supergravity in $d=8$ spacetime dimensions, let us extend the previous analysis so as to incorporate other groups into the game. Such groups typically appear as the U-dualities of the corresponding supergravity theory, a prototypical example being $SL(3, \mathbb{Z})$ (see e.g., \cite{Obers:1998fb}). The question then is: Can we analogously define automorphic forms for $SL(3, \mathbb{Z})$ following the same procedure as for the M\"obius group? The answer turns out to be yes, as we outline below.

Therefore, let us first introduce the appropriate elliptic $SL(3, \mathbb{Z})$-invariant operator, namely the Laplace operator on the coset space $SL(3, \mathbb{R})/SO(3)$
\beq \label{eq:laplacianSL3}
\Delta_3 = 4\tau_2^2 \partial_{\tau} \partial_{\bar \tau} + \frac{1}{\nu \tau_2} \left| \partial_b -\tau \partial_c \right|^2 + 3 \partial_{\nu} \left( \nu^2 \partial_{\nu}\right)\, ,
\eeq
where the precise physical meaning of the different coordinates will become clear later on in Appendix \ref{ap:maxSUGRA8d}. For the time being, it is enough to realise that the previous coordinates can be compactly grouped into the following $3\times3$ matrix (see e.g., \cite{Kiritsis:1997em})
\beq
 \mathcal{B}= \nu^{1/3} \begin{pmatrix}
		\frac{1}{\tau_2} \quad  \frac{\tau_1}{\tau_2} \quad \frac{c+\tau_1 b}{\tau_2}\\ \frac{\tau_1}{\tau_2} \quad  \frac{|\tau|^2}{\tau_2} \quad \frac{\tau_1 c+|\tau|^2 b}{\tau_2}\\ \frac{c+\tau_1 b}{\tau_2} \quad  \frac{\tau_1 c+|\tau|^2 b}{\tau_2} \quad \frac{1}{\nu} + \frac{|c+\tau b|^2}{\tau_2}
	\end{pmatrix}\, ,
\eeq
which moreover satisfies $\mathcal{B}=\mathcal{B}^{\text{T}}$ as well as $\det \mathcal{B}=1$. The reason why it is useful to introduce the matrix $\mathcal{B}$ is because it transforms in the adjoint representation of $SL(3, \mathbb{Z})$, namely upon performing some transformation $\mathcal{A} \in SL(3, \mathbb{Z})$, one finds that
\beq \label{eq:Btransf}
  \mathcal{B} \rightarrow \mathcal{A}^{\text{T}}\, \mathcal{B}\, \mathcal{A}\, .
\eeq
With this, we are now ready to define the Eisenstein $SL(3, \mathbb{Z})$ series of order $\ell$:
\beq\label{eq:SL3Eisenstein}
E_{\ell}^{sl_3} = \sum_{\mathbf{n}\, \in\, \mathbb{Z}^3 \setminus \lbrace \vec{0} \rbrace} \left( \sum_{i, j =1}^3 n_i\, \mathcal{B}^{ij}\, n_j\right)^{-\ell} = \sum_{\mathbf{n}\, \in\, \mathbb{Z}^3 \setminus \lbrace \vec{0} \rbrace} \nu^{-\ell/3} \left[ \frac{\left| n_1 + n_2 \tau + n_3 \left( c+\tau b\right)\right|^2}{\tau_2} + \frac{n_3^2}{\nu}\right]^{-\ell}\, ,
\eeq
with $\mathcal{B}^{ij}$ denoting the components of the inverse matrix of $\mathcal{B}$. Note that the above expression is manifestly $SL(3, \mathbb{Z})$-invariant, since the vector $\textbf{n} = \left( n_1, n_2, n_3\right)$ transforms as $\textbf{n} \rightarrow \mathcal{A}^{\text{T}}\, \textbf{n}$ under the duality group. As also happened with the non-holomorphic Eisenstein series defined in eq. \eqref{eq:nonholoEisenstein} above, the functions $E_{\ell}^{sl_3}$ are eigenvectors of the Laplacian $\Delta_3$, satisfying
\beq 
  \Delta_3 E_{\ell}^{sl_3} = \frac{2\ell (2\ell-3)}{3} E_{\ell}^{sl_3}\, .
\eeq
Let us also mention that the series $E_{\ell}^{sl_3}$, when viewed as a function of $\ell$, are absolutely convergent for $\ell>3/2$, whilst $E_{3/2}^{sl_3}$ is logarithmically divergent. This state of affairs reminds us of the situation for the modular non-holomorphic Eisenstein series $E_{\ell}^{sl_2}$, which had a simple pole for $\ell=1$. Therefore, proceeding analogously as in that case, one may define
\beq \label{eq:regularisation}
  \hat{E}_{3/2}^{sl_3} \equiv \lim_{\ell\to 3/2} \left( E_{\ell}^{sl_3} - \frac{2\pi}{\ell-3/2} - 4\pi(\gamma_{\text{e}}-1)\right)\, ,
\eeq
where again $\gamma_{\text{e}}$ denotes the Euler-Mascheroni constant. Such newly defined function is no longer singular and remains invariant under $SL(3, \mathbb{Z})$ transformations, with the price of not being a zero-mode of the Laplacian \eqref{eq:laplacianSL3} anymore.

\subsubsection*{Fourier-like expansions}

In what follows we will try to rewrite the $SL(3,\mathbb{Z})$ Eisenstein series in a way which makes manifest the perturbative and non-perturbative origin of the different terms that appear in the expansion, similarly to what we did for the $SL(2,\mathbb{Z})$ case. We closely follow Appendix A of \cite{Kiritsis:1997em}. First, let us introduce the following integral representation
\begin{align}\label{eq:integralrep}
	\notag E_{\ell}^{sl_3} &= \frac{\pi^\ell}{\Gamma(\ell)} \int_0^{\infty} \frac{dx}{x^{1+\ell}} \sum_{\mathbf{n}\, \in\, \mathbb{Z}^3 \setminus \lbrace \vec{0} \rbrace} \exp \left[-\frac{\pi}{x} \left( \sum_{i, j =1}^3 n_i\, \mathcal{B}^{ij}\, n_j\right) \right]\\
 &= \nu^{-\ell/3} \frac{\pi^\ell}{\Gamma(\ell)} \int_0^{\infty} \frac{dx}{x^{1+\ell}} \sum_{\mathbf{n}\, \in\, \mathbb{Z}^3 \setminus \lbrace \vec{0} \rbrace} \exp \left[-\frac{\pi}{x} \left( \frac{\left| n_1 + n_2 \tau + n_3 \left( c+ \tau b\right) \right|^2}{\tau_2} + \frac{n_3^2}{\nu}\right) \right]\, ,
\end{align}
which can be shown to coincide with the defining series \eqref{eq:SL3Eisenstein} after performing the change of variables $y=x^{-1}$ and using the definition of the $\Gamma$-function, namely 
\begin{align}
	\Gamma(z) = \int_0^{\infty} dy\, y^{z-1} e^{-y}\, .
\end{align}
After carefully separating the sum in the integers $n_i$ and performing a series of Poisson resummations (see footnote \ref{fnote:Poissonresummation}), one arrives at a Fourier series expansion of the form \cite{Kiritsis:1997em,Basu:2007ru,Basu:2007ck}
\begin{align}\label{eq:instexpSL3}
	\notag E_{\ell}^{sl_3} &= 2\nu^{-\ell/3} \tau_2^\ell \zeta(2\ell) + 2 \sqrt{\pi} T_2 \left( \tau_2 \nu^{1/3}\right)^{3/2-\ell} \frac{\Gamma(\ell-1/2)}{\Gamma(\ell)} \zeta(2\ell-1) + 2\pi \nu^{2\ell/3-1} \frac{\zeta(2\ell-2)}{\ell-1}\\
 &+  2 \frac{\pi^\ell \sqrt{\tau_2}}{\Gamma(\ell) \nu^{\ell/3}} \sum_{m,n \neq 0} \left| \frac{m}{n}\right|^{\ell-1/2} e^{2\pi \text{i} m n\tau_1}\, K_{\ell-1/2} (2\pi |m n| \tau_2)\, +\, \sum_{m, n \in \mathbb{Z} \setminus \lbrace (0,0) \rbrace} \mathcal{I}^\ell_{m, n}\, ,
\end{align}
where we have defined $T_2 \equiv \text{Im}\, T$, with $T= b+ \text{i} \left( \nu \tau_2\right)^{-1/2}$, and 
\begin{align}
	\mathcal{I}^\ell_{m, n} = 2\frac{\pi^\ell \nu^{\ell/6-1/2}}{\Gamma(\ell) \tau_2^{\ell/2-1/2}} \sum_{k \neq 0} \left| \frac{m+n\tau}{k}\right|^{\ell-1} e^{2\pi \text{i} k \left[n(c+\tau_1 b)- (m+n\tau_1)b \right]}\, K_{\ell-1} \left(2\pi |k|\frac{\left| m+n\tau \right|}{\sqrt{\nu \tau_2}}\right)\, .
\end{align}
Notice that upon using the Fourier expansion for the $SL(2,\mathbb{Z})$ series in eq. \eqref{eq:nonpertexpansion}, one can group the terms which depend on $\nu^{-\ell/3}$ into the following expression
\begin{align}\label{eq:SL3&SL2}
	 E_{\ell}^{sl_3} &= \nu^{-\ell/3} E_{\ell}^{sl_2}(\tau) + 2\pi \nu^{2\ell/3-1} \frac{\zeta(2\ell-2)}{\ell-1}\, +\, \sum_{m, n \in \mathbb{Z} \setminus \lbrace (0,0) \rbrace} \mathcal{I}^\ell_{m, n}\, .
\end{align}
The origin of each term in the above expansion will become clear when discussing maximal supergravity in eight spacetime dimensions, see Section \ref{ss:MthyT3}. For the moment, let us just say that the fact that the same modular Eisenstein series appear also in eq. \eqref{eq:SL3&SL2} can be understood from dimensional reduction arguments, since the 8d theory inherits certain perturbative and non-perturbative corrections from its 10d and 9d analogues, where the $SL(2,\mathbb{Z})$ duality group is all there is. Additionally, the sum over the pair of integers $(m,n)$ will be seen to correspond to certain instanton corrections arising from bound states of $(p,q)$-strings wrapping some $T^2$ of the internal geometry.

Furthermore, there exists another set of coordinates on $SL(3, \mathbb{R})/SO(3)$ apart from those employed in eq. \eqref{eq:laplacianSL3}, in which $\{\nu, \tau\}$ are exchanged with $\{\varphi_8, T\}$, where $e^{-2\varphi_8}=\tau_2^{3/2} \nu^{-1/2}$ and physically it corresponds to an eight-dimensional dilaton (see Section \ref{ss:MthyT3} and Appendix \ref{ap:maxSUGRA8d} for details). Using such parametrisation, one can expand $E_{\ell}^{sl_3}$ around `weak coupling' as follows\cite{Green:2010wi} 
\begin{align}\label{eq:instexpSL3-2}
	\notag E_{\ell}^{sl_3} &= 2 \zeta(2\ell) e^{-\frac{4\ell}{3} \varphi_8} + \pi^{1/2} \frac{\Gamma(\ell-1/2)}{\Gamma(\ell)} e^{-\left(\frac{2\ell}{3}-1 \right)\varphi_8} E_{\ell-1/2}^{sl_2} (T)\\
    \notag&+  \frac{2\pi^\ell}{\Gamma(\ell)} T_2^{\ell/2-1/4} e^{-\left(\frac{\ell}{3}-\frac{1}{2} \right)\varphi_8} \sum_{m,n \neq 0} \left| \frac{m}{n}\right|^{\ell-1/2} e^{2\pi \text{i} m n\tau_1}\, K_{\ell-1/2} (2\pi |m n| \tau_2)\\
    &+\frac{2\pi^\ell}{\Gamma(\ell)} T_2^{1/2} e^{\left(\frac{2\ell}{3}-1 \right)\varphi_8} \sum_{k \neq 0} \left| \frac{m+n\tau}{k}\right|^{\ell-1} e^{2\pi \text{i} k \left[n(c+\tau_1 b)- (m+n\tau_1)b \right]}\, K_{\ell-1} \left(2\pi |k| \left| m+n\tau \right| T_2\right)\, ,
\end{align}
where one should take $\tau_2$ as a function of $\lbrace\varphi_8, T_2\rbrace$ in the previous expression.

\subsubsection*{The $E_{3/2}^{sl_3}$ series}

To close this section, let us briefly discuss the particular case of the $SL(3,\mathbb{Z})$ Eisenstein series of order-$3/2$, since it plays a crucial role in our analysis in Section \ref{ss:MthyT3} from the main text. In fact, as already mentioned, $E_{\ell}^{sl_3}$, when seen as a function of the variable $\ell$, has a simple pole at $\ell=3/2$.\footnote{This is easy to see from eq. \eqref{eq:instexpSL3} above, since the functions $\zeta(1+x)$ as well as $\Gamma(x)$ present simple poles at $x=0$. Indeed, one obtains the following expansions around the pole:
\beq
\notag \zeta(1+\epsilon)= \frac{1}{\epsilon} + \gamma_{\text{e}} + \mathcal{O}(\epsilon)\, , \qquad \Gamma(\epsilon)=\frac{1}{\epsilon} - \gamma_{\text{e}} + \mathcal{O}(\epsilon)\, .
\eeq} Regularising in a way that preserves automorphicity (see eq. \eqref{eq:regularisation}), one finds for the series expansion the following expression \cite{Kiritsis:1997em}
\begin{align}\label{eq:Eisenstein3/2}
	\notag \hat{E}_{3/2}^{sl_3} &= 2\zeta(3) \frac{\tau_2^{3/2}}{\nu^{1/2}} + \frac{2 \pi^2}{3} T_2 + \frac{4\pi}{3} \log \nu\\
 &+  4\pi \sqrt{\frac{\tau_2}{\nu}} \sum_{m,n \neq 0} \left| \frac{m}{n}\right| e^{2\pi \text{i} m n\tau_1}\, K_{1} (2\pi |m n| \tau_2)\, +\, \sum_{m, n \in \mathbb{Z} \setminus \lbrace (0,0) \rbrace} \mathcal{I}^{3/2}_{m, n}\, ,
\end{align}
which in the limit \eqref{eq:instexpSL3-2} becomes \cite{Green:2010wi}
\begin{align}\label{eq:Eisenstein3/2-2}
	\hat{E}_{3/2}^{sl_3} &= 2\zeta(3) e^{-2\varphi_8} + 2 \hat{E}_{1}^{sl_2} (T) + \frac{4 \pi}{3} \varphi_8 + \mathcal{O} \left( \exp(-(T_2 e^{2\varphi_8})^{-1/2}),\exp(-(T_2^{-1} e^{2\varphi_8})^{-1/2})\right)\, .
\end{align}

\section{Maximal Supergravity in 8d}
\label{ap:maxSUGRA8d}

In this appendix we give some details regarding the construction of 8d $\mathcal{N}=2$ supergravity and the relation between the different relevant duality frames. The material presented here is complementary to the analysis of Section \ref{ss:MthyT3}. 

\subsection{The two-derivative action}
\label{ss:8daction}

Our strategy will be to first obtain the two-derivative bosonic action from M-theory compactified on $T^3$ in a two-step procedure, namely we first compactify on a two-dimensional torus, yielding maximal supergravity in 9d, and subsequently we reduce on a $S^1$. This allows us to relate the relevant physical quantities that appear in this setup, as studied in \cite{Calderon-Infante:2023ler}, with their Type II duals, where most of the higher-order corrections have been obtained.

Therefore, we proceed as in Section \ref{ss:MthyT2}, i.e. we compactify M-theory on $T^2$, yielding a 9d $\mathcal{N}=2$ theory with an action whose scalar-tensor sector reads
\begin{equation}\label{eq:9dap}
	S_\text{M-th}^{\text{9d}} = \frac{1}{2\kappa_9^2} \int d^{9}x\, \sqrt{-g}\,  \left[ R - \frac{9}{14} \frac{\left( \partial \mathcal{V}_2 \right)^2}{\mathcal{V}_2^2} -\frac{\partial \tau \cdot \partial \bar \tau}{2 \tau_2^2} \right]\, ,
\end{equation}
where $\mathcal{V}_2$ is the volume of the $T^2$ in 11d Planck units and $\tau=\tau_1+\text{i} \tau_2$ denotes its complex structure. Upon further compactifying on a circle of radius $R_3$ (in 9d Planck units), we obtain the following action
\begin{equation}\label{eq:8dap}
			\begin{aligned}
				S_\text{M-th}^{\text{8d}}\, &=\, \frac{1}{2\kappa_{8}^2} \int d^{8}x\sqrt{-g} \Bigg[R-\frac{9}{14} \left( \partial \log \mathcal{V}_2\right)^2 - \frac{7}{6} \left( \partial \log R_3\right)^2 -\frac{\partial \tau \cdot \partial \bar \tau}{2 \tau_2^2}\\
                &- \frac{\mathcal{V}_2^{-12/7} R_3^{-2}}{2} \left( \partial C_{123}^{(3)}\right)^2 -\frac{\mathcal{V}_2^{9/7} R_3^{-2}}{2 \tau_2} \left| \partial \left( \frac{\text{Im}\, (\tau \bar{\xi}_{\text{M}})}{\tau_2}\right) + \tau\, \partial \left( \frac{\text{Im}\, (\xi_{\text{M}})}{\tau_2}\right)\right|^2\Bigg]\, ,
			\end{aligned}
\end{equation}
where $C_{123}^{(3)}$ arises from the reduction of the 11d 3-form $C_3$ along the $T^3$, whilst $\xi_{\text{M}}=-C_1^{1}+\text{i} C_2^{1} \tau_2$ with $\lbrace C_1^{1},C_2^{1} \rbrace$ being compact scalar fields parametrising the orientation of the two-dimensional torus within the overall $T^3$.

Several comments are now in order. First, notice that one can relate the moduli fields appearing in eq. \eqref{eq:8dap} with their canonically normalised analogues, as defined in \cite{Calderon-Infante:2023ler}. In particular, for the non-compact scalars one finds
\begin{equation} \label{eq:8dchangeofvariables}
	\begin{split} 
		\hat \rho &= \sqrt{\frac{7}{6}} \log R_3\, , \qquad \hat U = \frac{3}{\sqrt{14}} \log \mathcal{V}_2 \, ,\\
		\hat \tau &= \frac{1}{\sqrt{2}} \log \tau_2  \, .
	\end{split}
\end{equation}
Second, let us also mention that one can rewrite the action \eqref{eq:8dap} above in a way that makes manifest the $SL(2,\mathbb{R})\times SL(3,\mathbb{R})$ symmetry that the theory enjoys at the \emph{classical} level.\footnote{Quantum-mechanically, the symmetry group is broken due to instanton effects down to its discrete $SL(2,\mathbb{Z})\times SL(3,\mathbb{Z})$ part, see Section \ref{ss:higherderivative} below.} This requires us to extract the overall $T^3$ volume, which is given by
\begin{equation} \label{eq:T3volume}
     \mathcal{V}_3 = \mathcal{V}_2\, R_3\, \frac{\ell_9}{\ell_{11}} = \mathcal{V}_2^{6/7}\, R_3\, ,
\end{equation}
where the 9d and 11d Planck lengths are related by $\ell_{11}^7=\ell_9^7\, \mathcal{V}_2$, as well as repackage the action as follows
\begin{align}\label{eq:8dsl3}
	S_\text{M-th}^{\text{8d}}\, =\, &\frac{1}{2\kappa_8^2} \int d^{8}x\, \sqrt{-g}\,  \left[ R + \frac{1}{4} \text{tr} \left( \partial \tilde{g} \cdot \partial \tilde{g}^{-1} \right) -\frac{\partial \mathcal{T} \cdot \partial \bar{\mathcal{T}}}{2 \mathcal{T}_2^2} \right]\, ,
\end{align}
where the $3\times3$ matrix $\tilde{g}$ is obtained from the internal metric of the three-dimensional torus with the overall volume extracted, i.e. $\tilde{g}_{ij}= \mathcal{V}_3^{-1/3} g_{ij}$, and we have also defined the complex field $\mathcal{T}=C_{123}^{(3)} + \text{i} \mathcal{V}_3$ in the previous expression. Note that the $SL(2,\mathbb{R})$ group acts on the variable $\mathcal{T}$, whilst the $SL(3,\mathbb{R})$ transformations only affect the matrix $\tilde g$, which moreover transforms in the adjoint (see discussion around eq. \eqref{eq:Btransf}).

With this, we can now perform several duality transformations so as to match the action functional in eq. \eqref{eq:8dap} with the corresponding ones arising from compactifying Type II String Theory on $T^2$. To do so, we first relate M-theory with Type IIA String Theory, and then upon T-dualising, we translate the result into the Type IIB dual description. Therefore, upon taking the circle with radius $R_3$ to be indeed the M-theory circle, one arrives at Type IIA String Theory, where the following identifications should be made
\begin{equation} \label{eq:8dMthy/IIA}
	\begin{split} 
		\mathcal{T} &\longleftrightarrow T=b+\text{i} T_2\, ,\\
            \tau &\longleftrightarrow U=U_1+\text{i} U_2\, ,\\
		\mathcal{V}_2^{9/7}\, R_3^{-2} &\longleftrightarrow e^{-2\varphi_8}=e^{-2\phi}\, T_2\, ,
	\end{split}
\end{equation}
where the LHS variables correspond to the M-theory ones, whilst the RHS denote the Type IIA moduli. These are the usual (complexified) K\"ahler modulus $T$, whose imaginary part controls the volume of the Type IIA torus in string units, $U$ denotes its complex structure and $\varphi_8$ is the 8d dilaton. Hence, after performing the change of variables in \eqref{eq:8dMthy/IIA} one arrives at an action of the form
\begin{equation}\label{eq:8dIIA}
			\begin{aligned}
				S_\text{IIA}^{\text{8d}}\, &=\, \frac{1}{2\kappa_{8}^2} \int d^{8}x\sqrt{-g} \Bigg[R - \frac{2}{3} \left( \partial \varphi_8\right)^2-\frac{\partial T \cdot \partial \bar T}{2 T_2^2}  -\frac{\partial U \cdot \partial \bar U}{2 U_2^2}\\
                & -\frac{e^{-2\varphi_8}}{2 U_2} \left| \partial \left( \frac{\text{Im}\, (U \bar{\xi}_{\text{A}})}{U_2}\right) + U\, \partial \left( \frac{\text{Im}\, (\xi_{\text{A}})}{U_2}\right)\right|^2\Bigg]\, ,
			\end{aligned}
\end{equation}
where $\xi_{\text{A}}=-C_1^{1}+\text{i} C_2^{1} U_2$ with $\lbrace C_1^{1},C_2^{1} \rbrace$ being interpreted now as coming from the reduction of the RR 1-form $C_1$ on any of the two one-cycles within the $T^2$.

On a next step, we perform some T-duality which relates the two Type II string theories in eight dimensions. Therefore, the familiar Buscher rules translate into the exchange of K\"ahler and complex structure moduli, i.e. $T \leftrightarrow U$, whilst the 8d dilaton is left unchanged. Additionally, one finds for the complex $\xi_{\text{B}}$ field the following new expression
\begin{equation}
    \xi_{\text{B}}= -b+\text{i} \tau_1\, T_2\, ,
\end{equation}
where $\tau_1=C_0$ corresponds to the RR 0-form of Type IIB String Theory. This yields an action for the scalar-tensor sector of the theory of the form
\begin{equation}\label{eq:8dIIB}
			\begin{aligned}
				S_\text{IIB}^{\text{8d}}\, &=\, \frac{1}{2\kappa_{8}^2} \int d^{8}x\sqrt{-g} \Bigg[R - \frac{2}{3} \left( \partial \varphi_8\right)^2 -\frac{\partial U \cdot \partial \bar U}{2 U_2^2}-\frac{\partial T \cdot \partial \bar T}{2 T_2^2}\\
                & -\frac{e^{-2\varphi_8}}{2 T_2} \left| \partial \left( \frac{\text{Im}\, (T \bar{\xi}_{\text{B}})}{T_2}\right) + T\, \partial \left( \frac{\text{Im}\, (\xi_{\text{B}})}{T_2}\right)\right|^2\Bigg]\, .
			\end{aligned}
\end{equation}

Let us also mention that, even though the Type IIB action only exhibits $SL(2,\mathbb{Z})_T \times SL(2,\mathbb{Z})_U$ invariance in eq. \eqref{eq:8dIIB}, the full 8d theory is known to accommodate a larger $SL(2,\mathbb{Z})\times SL(3,\mathbb{Z})$ duality group, as discussed around eq. \eqref{eq:8dsl3}. This latter fact can be made manifest upon performing a change of (moduli) coordinates of the form $\lbrace \varphi_8, T\rbrace \to \lbrace \nu, \tau\rbrace$, where $\nu=\left( \tau_2\, T_2^2 \right)^{-1}$ and $\tau$ is the Type IIB axio-dilaton. Upon doing so, one arrives at 
\begin{equation}\label{eq:IIB8d-2}
			\begin{aligned}
				S_\text{IIB}^{\text{8d}}\, =\, & \frac{1}{2\kappa_{8}^2} \int d^{8}x\sqrt{-g} \left[R-\frac{1}{6}\frac{(\partial \nu)^2}{\nu^2} -\frac{\partial \tau \cdot \partial \bar \tau}{2 \tau_2^2} -\frac{\partial U \cdot \partial \bar U}{2 U_2^2} - \nu \frac{\left| \tau \partial b + \partial c\right|^2}{2\tau_2}\right]\, .
			\end{aligned}
\end{equation}
Notice that now the action becomes manifestly invariant under the modular symmetry $SL(2,\mathbb{Z})_{\tau} \times SL(2,\mathbb{Z})_U$, where the first factor acts on the axio-dilaton and is different from the one transforming the K\"ahler structure. In fact, as explained in Section \ref{ss:MthyT3} one can introduce a $3\times3$ matrix with unit determinant, $\mathcal{B}$, as follows\cite{Liu:1997mb}
\beq\label{eq:SL3matrixap}
 \mathcal{B}= \nu^{1/3} \begin{pmatrix}
		\frac{1}{\tau_2} \quad  \frac{\tau_1}{\tau_2} \quad \frac{c+\tau_1 b}{\tau_2}\\ \frac{\tau_1}{\tau_2} \quad  \frac{|\tau|^2}{\tau_2} \quad \frac{\tau_1 c+|\tau|^2 b}{\tau_2}\\ \frac{c+\tau_1 b}{\tau_2} \quad  \frac{\tau_1 c+|\tau|^2 b}{\tau_2} \quad \frac{1}{\nu} + \frac{|c+\tau b|^2}{\tau_2}
	\end{pmatrix}\, ,
\eeq
in terms of which the previous 8d action reads as (c.f. eq. \eqref{eq:8dsl3})
\begin{equation}\label{eq:IIB8dSL3ap}
			\begin{aligned}
				S_\text{IIB}^{\text{8d}}\, =\, & \frac{1}{2\kappa_{8}^2} \int d^{8}x\sqrt{-g} \left[R -\frac{\partial U \cdot \partial \bar U}{2 U_2^2} + \frac{1}{4} \text{tr} \left( \partial \mathcal{B} \cdot \partial \mathcal{B}^{-1} \right) \right]\, ,
			\end{aligned}
\end{equation}
thus exhibiting the full duality symmetry of the theory.

\subsection{Higher-derivative corrections}
\label{ss:higherderivative}

Our discussion so far has been restricted to the two-derivative action, where quantum effects do not play any role. However, when considering certain physical processes, such as four-point graviton scattering, one has to take into account several effects of quantum-mechanical origin, which are typically encapsulated by various higher-dimensional operators $\mathcal{O}_n(R)$ in the effective action. In fact, according to our general discussion in Section \ref{s:speciesscale}, one expects such \emph{gravitational} EFT series to break down precisely at the QG cut-off $\LQG$, namely the species scale. Therefore, it is of great interest to study those operators as well as their moduli dependence.

The first non-trivial correction to the bosonic action \eqref{eq:IIB8d-2} includes an operator comprised by four Weyl tensors contracted in a particular way. This has been computed in the past using a variety of methods, ranging from one-loop calculations in M-theory to non-perturbative instanton computations. The result is \cite{Green:1997as,Green:2005ba} (see also \cite{Kiritsis:1997em,Basu:2007ru,Basu:2007ck})
\beq
S_{R^4,\, \text{IIB}}^{\text{8d}}= \int d^{8}x \sqrt{-g}\, \left( \hat{E}_{3/2}^{sl_3} +2\hat{E}_{1}^{sl_2}\right) t_8 t_8 R^4\, ,
\label{eq:8dR^4IIB}
\eeq
where $t_8 t_8 R^4$ denotes a particular contraction of four Riemann tensors \cite{Green:1981ya}. The functions $\hat{E}_{3/2}^{sl_3}$ and $\hat{E}_{1}^{sl_2}$ are (appropriately regularised) Eisenstein series of order $3/2$ and 1 for the duality groups $SL(3,\mathbb{Z})$ and $SL(2,\mathbb{Z})$, respectively. They can be expanded as follows (see Section \ref{ap:Massform}):
\begin{align}\label{eq:instexpSL3ap}
	\notag \hat{E}_{3/2}^{sl_3} &= 2\zeta(3) \frac{\tau_2^{3/2}}{\nu^{1/2}} + \frac{2 \pi^2}{3} T_2 + \frac{4\pi}{3} \log \nu\\
 &+  4\pi \sqrt{\frac{\tau_2}{\nu}} \sum_{m,n \neq 0} \left| \frac{m}{n}\right| e^{2\pi \text{i} m n\tau_1}\, K_{1} (2\pi |m n| \tau_2)\, +\, \sum_{m, n \in \mathbb{Z} \setminus \lbrace (0,0) \rbrace} \mathcal{I}^{3/2}_{m, n}\, ,
\end{align}
with
\begin{align}\label{eq:I3/2mn}
	\mathcal{I}^{3/2}_{m, n} = 2\frac{\pi^{3/2} \nu^{-1/4}}{\Gamma({3/2}) \tau_2^{1/4}} \sum_{k \neq 0} \left| \frac{m+n\tau}{k}\right|^{1/2} e^{2\pi \text{i} k \left[n(c+\tau_1 b)- (m+n\tau_1)b \right]}\, K_{1/2} \left(2\pi |k|\frac{\left| m+n\tau \right|}{\sqrt{\nu \tau_2}}\right)\, ,
\end{align}
for the $SL(3,\mathbb{Z})$-invariant piece, whilst the second factor in eq. \eqref{eq:8dR^4IIB} reads as
\beq
2\hat{E}_{1}^{sl_2}\, =\, -2\pi \text{log} \left(U_2\,|\eta(U)|^4\right)\, .
\eeq
The physical origin of both terms can be easily understood upon looking at the variables entering in each of the two expansions. Indeed, $\hat{E}_{1}^{sl_2}$ encodes certain KK threshold corrections, which depend on the complex structure of the torus, whilst the structure of $\hat{E}_{\ell}^{sl_3}$ is richer: It contains both $\alpha'$ and $g_s$ \emph{perturbative} contributions, together with the D$(-1)$-instanton series --- which is already present in 10d, see Section \ref{ss:10dIIB} --- as well as non-perturbative $(p,q)$-string instantons, whose action is controlled by quantity entering the modified Bessel function in \eqref{eq:I3/2mn}, namely $\frac{\left| m+n\tau \right|}{\sqrt{\nu \tau_2}}=\left| m+n\tau \right|T_2$.

Before proceeding any further, let us note that if instead of the $\lbrace \nu, \tau\rbrace$-parametrisation one chooses the alternative $\lbrace \varphi_8, T\rbrace$ coordinates, whose leading order action is shown in eq. \eqref{eq:8dIIB}, the $SL(3,\mathbb{Z})$-series can be expanded as follows
\begin{align}\label{eq:Eisenstein3/2-2ap}
	\hat{E}_{3/2}^{sl_3} &= 2\zeta(3) e^{-2\varphi_8} + 2 \hat{E}_{1}^{sl_2} (T) + \frac{4 \pi}{3} \varphi_8 + \mathcal{O} \left( \exp(-e^{-\varphi_8})\right)\, ,
\end{align}
which of course agrees with the first few terms of \eqref{eq:instexpSL3ap}.

In order to connect with our discussion in the M-theory picture \cite{Calderon-Infante:2023ler}, we need to rewrite the previous $R^4$ correction in terms of the variables adapted to the action in eq. \eqref{eq:8dap}. To do so, it will be enough to know how the 8d dilaton $\varphi_8$ and K\"ahler modulus $T_2$ depend on the M-theory variables $\lbrace \mathcal{V}_2, R_3, \tau_2\rbrace$. Thus, upon using the map between Type IIB and M-theory descriptions
\begin{equation} \label{eq:8dMthy/IIB}
	\begin{split} 
		\mathcal{T} &\longleftrightarrow U\, ,\\
            \tau &\longleftrightarrow T\, ,\\
		\mathcal{V}_2^{9/7}\, R_3^{-2} &\longleftrightarrow e^{-2\varphi_8}\, ,
	\end{split}
\end{equation}
as well as the expression of the $T^3$ volume in terms of $\mathcal{V}_2$ and $R_3$ (see eq. \eqref{eq:T3volume}), one finds
\begin{equation} \label{eq:usefulmap}
      \mathcal{V}_2 = T_2^{\frac{2}{3}}\, e^{-\frac{2\varphi_8}{3}}\, , \qquad R_3 = T_2^{\frac{3}{7}}\, e^{\frac{4\varphi_8}{7}}\, .
\end{equation}
From these one may even obtain the expression of the Type IIB coordinate $\nu$ in terms of M-theory variables, which reads $\nu=\mathcal{V}_2^{-18/7}\, R_3^4\, \tau_2^{-3/2}$.

Using the previous expressions one may rewrite the first few terms of $\hat{E}_{3/2}^{sl_3}$ as
\begin{align}\label{eq:instexpSL3Mth1}
	\hat{E}_{3/2}^{sl_3} &= 2\zeta(3) \mathcal{V}_2^{9/7}\, R_3^{-2} + \frac{2\pi^2}{3} \tau_2 - 2\pi \text{log} (\tau_2) - \frac{2 \pi}{3} \log \left( \mathcal{V}_2^{9/7}\, R_3^{-2}\right) + \ldots\, ,
\end{align}
where the ellipsis indicates further non-perturbative contributions.
    
\end{appendix}

\bibliography{refs-modular}
\bibliographystyle{JHEP}

\end{document}